\documentclass[5p,twocolumn]{elst}
\usepackage{graphicx,color,xcolor}
\usepackage{xcolor} 
\usepackage{array} 

\usepackage{amsmath,amsfonts,amssymb,bm}
\usepackage{algorithm,algorithmicx,algpseudocode}
\algnewcommand{\LineComment}[1]{\Statex \(\triangleright\) #1}    
\usepackage{accents}
\newcommand{\ubar}[1]{\underaccent{\bar}{#1}}
\DeclareMathOperator*{\argmin}{arg\,min}
\DeclareMathOperator*{\diag}{\textit{diag}}
\DeclareMathOperator*{\trace}{tr}
\DeclareMathOperator*{\rank}{rank}
\definecolor{Blue}{RGB}{88, 105, 225}
\definecolor{dred}{rgb}{0.6,0,0}

\def\va{{\bm{a}}}
\def\vb{{\bm{b}}}
\def\vc{{\bm{c}}}
\def\vd{{\bm{d}}}

\def\vu{{\bm{u}}}
\def\vv{{\bm{v}}}
\def\vw{{\bm{w}}}
\def\vx{{\bm{x}}}
\def\vy{{\bm{y}}}

\def\mA{{\bm{A}}}
\def\mB{{\bm{B}}}
\def\mC{{\bm{C}}}

\def\mG{{\bm{G}}}
\def\mH{{\bm{H}}}
\def\mI{{\bm{I}}}

\def\mL{{\bm{L}}}
\def\mM{{\bm{M}}}

\def\mP{{\bm{P}}}
\def\mQ{{\bm{Q}}}
\def\mR{{\bm{R}}}
\def\mS{{\bm{S}}}

\def\mW{{\bm{W}}}

\def\mZ{{\bm{Z}}}

\newtheorem{thm}{Theorem}

\newtheorem{lem}[thm]{Lemma}
\newtheorem{claim}[thm]{Claim}

\newtheorem{assum}[thm]{Assumption}
\newtheorem{prop}[thm]{Proposition}

\newtheorem{rem}[thm]{Remark}

\usepackage{epstopdf}
\usepackage{natbib}

\journal{}
\begin{document}
\begin{frontmatter}
\title{Constrained Attack-Resilient Estimation of Stochastic Cyber-Physical Systems}

\author[1]{Wenbin Wan\corref{cor1}}      \ead{wenbinw2@illinois.edu}
\author[2]{Hunmin Kim}                    \ead{kim\_h@mercer.edu}
\author[1]{Naira Hovakimyan}              \ead{nhovakim@illinois.edu}
\author[3]{Petros Voulgaris}              \ead{pvoulgaris@unr.edu}

\cortext[cor1]{Corresponding author.}

\address[1]{University of Illinois at Urbana-Champaign, USA}  
\address[2]{Mercer University, USA} 
\address[3]{University of Nevada, Reno, USA} 
\begin{abstract}
In this paper, a constrained attack-resilient estimation algorithm (\texttt{CARE}) is developed for stochastic cyber-physical systems. The proposed \texttt{CARE} can simultaneously estimate the compromised system states and attack signals. It has improved estimation accuracy and attack detection performance when physical constraints and operational limitations are available. In particular, \texttt{CARE} is designed for simultaneous input and state estimation that provides minimum-variance unbiased estimates, and these estimates are projected onto the constrained space restricted by inequality constraints subsequently. We prove that the estimation errors and their covariances from \texttt{CARE} are less than those from unconstrained algorithms and confirm that this property can further reduce the false negative rate in attack detection. We show that estimation errors of \texttt{CARE} are practically exponentially stable in mean square. Finally, an illustrative example of attacks on a vehicle is given to demonstrate the improved estimation accuracy and detection performance compared to an existing unconstrained algorithm.
\end{abstract}
\end{frontmatter}

\section{Introduction}
Cyber-Physical Systems (CPS) play a vital role in the metabolism of applications from large-scale industrial systems to critical infrastructures, such as smart grids, transportation networks, precision agriculture, and industrial control systems~\cite{rajkumar2010cyber}. Recent developments in CPS and their safety-critical applications have led to a renewed interest in CPS security. The interaction between information technology and the physical system has made control components of CPS vulnerable to malicious attacks~\cite{cardenas2008research}. Recent cases of CPS attacks have clearly illustrated their susceptibility and raised awareness of the security challenges in these systems. These include attacks on large-scale critical infrastructures, such as the German steel mill cyber attack~\cite{lee2014german}, and Maroochy Water breach~\cite{slay2007lessons}. Similarly, malicious attacks on avionics and automotive vehicles have been reported, such as the U.S. drone RQ-170 captured in Iran~\cite{peterson2011iran}, and disabling the brakes and stopping the engine on civilian cars~\cite{koscher2010experimental,checkoway2011comprehensive}.

\paragraph{Related work}
Traditionally, most works in the field of attack detection had only focused on monitoring the cyber-space misbehavior~\cite{raiyn2014survey}. With the emergence of CPS, it becomes vitally important to monitor physical misbehavior as well because the impact of the attack on physical systems also needs to be addressed~\cite{cardenas2008secure}. In the last decade, attention has been drawn from the perspective of the control theory that exploits some prior information on the system dynamics for detection and attack-resilient control. For instance, a unified modeling framework for CPS and attacks is proposed in~\cite{pasqualetti2013attack}. A typical control architecture for the networked system under both cyber and physical attacks is proposed in~\cite{teixeira2012attack}; then attack scenarios, such as Denial-of-Service (DoS) and false-data injection (FDI) are analyzed using this control architecture in~\cite{teixeira2015secure}.

In recent years, model-based detection has been tremendously studied. Attack detection has been formulated as an $\ell_0$/$\ell_\infty$ optimization problem, which is NP-hard~\cite{pajic2014robustness}. A convex relaxation has been studied in~\cite{fawzi2014secure}. Furthermore, the worst-case estimation error has been analyzed in~\cite{pajic2017attack}. 
Multirate sampled data controllers have been studied to guarantee detectability in~\cite{NAGHNAEIAN201912} and to detect zero-dynamic attacks in~\cite{8796181}. A residual-based detector has been designed for power systems against false-data injection attacks, and the impact of attacks has been analyzed in~\cite{liu2011false}. 

In addition, some papers have studied active detection, such as~\cite{mo2009secure,mo2014detecting}, where the control input is watermarked with a pre-designed scheme that sacrifices optimality. The aforementioned methods have the problem that the state estimate is not resilient concerning the attack signal, and incorrect state estimates make it more challenging for defenders to react to malicious attacks consequently.

Attack-resilient estimation and detection problems have been studied to address the above challenge in~\cite{yong2015resilients,forti2016bayesian,kim2020simultaneous}, where attack detection has been formulated as a simultaneous input and state estimation problem, and the minimum-variance unbiased estimation technique has been applied. More specifically, the approach has been applied to linear stochastic systems in~\cite{yong2015resilients}, stochastic random set methods in~\cite{forti2016bayesian}, and nonlinear systems in~\cite{kim2020simultaneous}. These detection algorithms rely on statistical thresholds, such as the $\chi^2$ test, which is widely used in attack detection~\cite{mo2014detecting,teixeira2010cyber}. Since the detection accuracy improves when the covariance decreases, a smaller covariance is desired.

On top of the minimum-variance estimation approach, the covariance can be further reduced when we incorporate the information of the input and state in terms of constraints. There have been several investigations on Kalman filtering with state constraints~\cite{simon2002kalman,ko2007state,simon2010kalman,kong2021kalman}. The state constraints are induced by unmodeled dynamics and operational processes. Some of these examples include vision-aided inertial navigation~\cite{mourikis2007multi}, target tracking~\cite{wang2002filtering} and power systems~\cite{yong2015simultaneous}. Constraints on inputs are also considered, such as avoiding reachable dangerous states under the assumption that the attack input is constrained~\cite{kafash2018constraining} and designing a resilient controller based on the partial knowledge of the attacker in terms of inequality constraints~\cite{djouadi2015finite}. The methods in~\cite{kafash2018constraining,djouadi2015finite} can efficiently be used to maneuver a class of attacks when input inequality constraints are available but cannot resiliently address the estimation problem due to the false-data injection. This problem remains to be solved with a stability guarantee in the presence of inequality constraints. In the current paper, we aim to solve the resilient estimation problem and investigate the stability and performance of the algorithm design that integrates with information aggregation. To the best of our knowledge, this is the first investigation that considers both state and input inequality constraints for attack-resilient estimation with guaranteed stability.

\paragraph{Contributions}
Our main contributions of this work can be summarized as follows.
    \textit{i)} We propose a constrained attack-resilient estimation algorithm (\texttt{CARE}) that can estimate the compromised system states and the attack signals simultaneously. \texttt{CARE} first provides minimum-variance unbiased estimates, and then they are projected onto the constrained space induced by information aggregation.
    \textit{ii)} The proposed \texttt{CARE} has better estimation performance. We show that the projection strictly reduces the estimation errors and covariances.
    \textit{iii)} We are the first to investigate the stability of the estimation algorithm with inequality constraints and prove that the estimation errors are practically exponentially stable in mean square.
    \textit{iv)} The proposed \texttt{CARE} has better attack detection performance. We provide rigorous analysis that the false negative rate is reduced by using the proposed algorithm.
    \textit{v)} The proposed algorithm is compared with the state-of-the-art method to show the improved estimation and attack detection performance.

\paragraph{Paper organization}
The rest of the paper is organized as follows. Section~\ref{pre} introduces notations, $\chi^2$ test for detection, and problem formulation. The high-level idea of the proposed algorithm is presented in Section \ref{sec:alsop}. Section \ref{algodetail} gives a detailed algorithm derivation. Section \ref{analysis} demonstrates the performance improvement and investigates the stability analysis of the proposed algorithm. Section \ref{simulation} presents an illustrative example of vehicle attacks. Finally, Section~\ref{conclusion} draws the conclusion.

\section{Preliminaries} \label{pre}

\paragraph{Notations}
We use the subscript $k$ to denote the time index. For a real set ${\mathbb R}$, ${\mathbb R}^n_+$ denotes the set of positive elements in the $n$-dimensional Euclidean space and ${\mathbb R}^{n \times m}$ denotes the set of all $n \times m$ real matrices. For a matrix $\mA$, $\mA^\top$, $\mA^{-1}$, $\mA^\dagger$, $\diag(\mA)$, $\trace(\mA)$ and $\rank(\mA)$ denote the transpose, inverse, Moore-Penrose pseudoinverse, diagonal, trace and rank of $\mA$, respectively. For a symmetric matrix $\mS$, $\mS > 0$ ($\mS \geq 0$) indicates that $\mS$ is positive (semi)definite. The matrix $\mI$ denotes the identity matrix with an appropriate dimension. We use $\|\cdot\|$ to denote the standard Euclidean norm for vector or an induced matrix norm if it is not specified, ${\mathbb E}[\,\cdot\,]$ to denote the expectation operator, and $\times$ to denote matrix multiplication when the multiplied terms are in different lines. For a vector $\va$, $\va(i)$ denotes the $i^{th}$ element in the vector $\va$.  Finally, vectors $\va$, $\hat{\va}$, $\tilde{\va} \triangleq \va - \hat{\va}$ denote the ground truth, estimate and estimation error of $\va$, respectively. 

\paragraph{$\chi^2$ Test for Detection}
Given a sample of Gaussian random variable $\hat{\bm{\sigma}}_k$  with unknown mean $\bm{\sigma}_k$ and known covariance $\bm{\Sigma}_k$, the $\chi^2$ test provides statistical evidence of whether $\bm{\sigma}_k=0$ or not. In particular, the sample $\hat{\bm{\sigma}}_k$ is being  normalized by $\hat{\bm{\sigma}}_k^\top \bm{\Sigma}_k^{-1}\hat{\bm{\sigma}}_k$, and we compare the normalized value with $\chi_{df}^2(\alpha)$, where $\chi_{df}^2(\alpha)$ is the $\chi^2$ value with degree of freedom $df$ and statistical significance level $\alpha$. We reject the null hypothesis $H_0$: $\bm{\sigma}_k=0$, if $\hat{\bm{\sigma}}_k^\top \bm{\Sigma}_k^{-1}\hat{\bm{\sigma}}_k > \chi^2_{df}(\alpha)$, and accept alternative hypothesis $H_1$: $\bm{\sigma}_k \neq 0$, i.e., there is significant statistical evidence that $\bm{\sigma}_k$ is non-zero. Otherwise, we accept $H_0$, i.e., there is no significant evidence that $\bm{\sigma}_k$ is non-zero.

\paragraph{False negative rate}
Given a set of vectors $\{\bm{\sigma}_k\}$, the false negative rate of the $\chi^2$ test is defined as the ratio of the number of false negative test results $N_{neg}$ and the number of non-zero vectors in the given set $N_{\bm{\sigma}_k \neq 0}$
\begin{align}
F_{neg}(\{\hat{\bm{\sigma}}_k\}, \{\bm{\Sigma}_k\})& \triangleq \frac{N_{neg}}{N_{\bm{\sigma}_k \neq 0}} = \frac{\sum_{k} (\bm{1}_k)}{N_{\bm{\sigma}_k \neq 0}}, \label{eq: Fneg define}
\end{align}
where
\begin{align} \label{eq: indicator}
\bm{1}_k \triangleq
\begin{cases}
1, & \textit{if } \hat{\bm{\sigma}}_k^\top \bm{\Sigma}_k^{-1}\hat{\bm{\sigma}}_k \leq \chi^2_{df}(\alpha) \textit{ and } \bm{\sigma}_k \neq 0 \\
0, & \textit{otherwise}
\end{cases}.
\end{align}

\paragraph{Problem Formulation}
Consider the following linear time-varying (LTV) discrete-time stochastic system\footnote{ The current paper considers a general formulation for the attack input matrix $\mG_{k}$. If $\vd_k$ is injected into the control input, then  $\mG_k = \mB_k$. If $\vd_k$ is directly injected into the system, then $\mG_k = \mI$.}
\begin{subequations}\label{eq: sys}
\begin{align}
  \vx_{k+1} &= \mA_k \vx_k + \mB_k \vu_k+ \mG_{k} \vd_k + \vw_k \label{eq: sys dyna}\\
     \vy_{k} &= \mC_{k}\vx_k + \vv_{k}, \label{eq: sys measure}
\end{align}
\end{subequations}
where $\vx_k \in \mathbb{R}^m$, $\vu_k \in \mathbb{R}^n$ and $\vy_{k}\in \mathbb{R}^{n_y}$ are the state, the control input and the sensor measurement, respectively. The attack signal is modeled as a simultaneous input $\vd_k \in \mathbb{R}^{n_d}$, which is unknown to the defender. System matrices $\mA_k$, $\mB_k$, $\mC_k$ and $\mG_k$ are known and bounded with appropriate dimensions. We assume that $\rank(\mC_{k} \mG_{k-1}) = n_d$, $0 \leq n_d \leq n_y$. This is typical assumption as in~\cite{yong2016unified,kim2017attack}. The interpretation of this assumption is that the impact of the attack $\vd_{k-1}$ on the system dynamics can be observed by $\vy_k$. The process noise $\vw_k$ and the measurement noise $\vv_k$ are assumed to be i.i.d. Gaussian random variables with zero means and covariances $\mQ_k \triangleq \mathbb{E}[\vw_k \vw_k^\top] \geq 0$ and $\mR_k \triangleq \mathbb{E}[\vv_k \vv_k^\top] > 0$. Moreover, the measurement noise $\vv_k$, the process noise $\vw_k$, and the initial state $\vx_0$ are uncorrelated with each other.

The adopted attack model in~\eqref{eq: sys} is known as the FDI attack that is a very general type of attack, and includes physical attacks, Trojans, replay attacks, overflow bugs, packet injection, etc~\cite{guo2018roboads}. Because of this generality, this attack model has been widely used in  CPS security literature (e.g.,~\cite{pasqualetti2013attack,teixeira2015secure,yong2015resilients}).

In the cyber-space, digital attack signals could be unconstrained, but their impact on the physical world is restricted by physical and operational constraints (i.e., $\vx_k$ and $\vd_k$ are constrained). For example, a vehicle has a limit on acceleration, velocity, steering angle, and change of steering angle. Any physical constraints and ability limitations on attack signals and states are presented by the inequality constraints
\begin{align}\label{eq: const}
    \mathcal{A}_k\vd_k \leq \vb_k, \ \mathcal{B}_k \vx_k \leq \vc_k,
\end{align}
where matrices $\mathcal{A}_k$, $\mathcal{B}_k$, and vectors $\vb_k$, $\vc_k$ are known and bounded with appropriate dimensions. 
Throughout this paper, we assume that the feasible sets of the constraints in~\eqref{eq: const} are non-empty.

\begin{rem}
Gaussian noise in~\eqref{eq: sys} is one of the general ways to model physical systems so that the filtering algorithms use this model to track the level of uncertainties. Therefore, many pieces of work consider Gaussian noise even in the presence of bounded constraints~\cite{teixeira2009state,simon2002kalman,simon2010kalman}.
\end{rem}

\paragraph{Problem statement}
Given the stochastic system in~\eqref{eq: sys}, we aim to design an attack-resilient estimation algorithm that can simultaneously estimate the compromised system state $\vx_k$ and the attack signal $\vd_k$. In addition, we seek to improve estimation accuracy and detection performance with a stability guarantee when incorporating the information of the input and state in terms of constraints in~\eqref{eq: const}.

\begin{figure}[ht]
\begin{center}
\includegraphics[width=6cm]{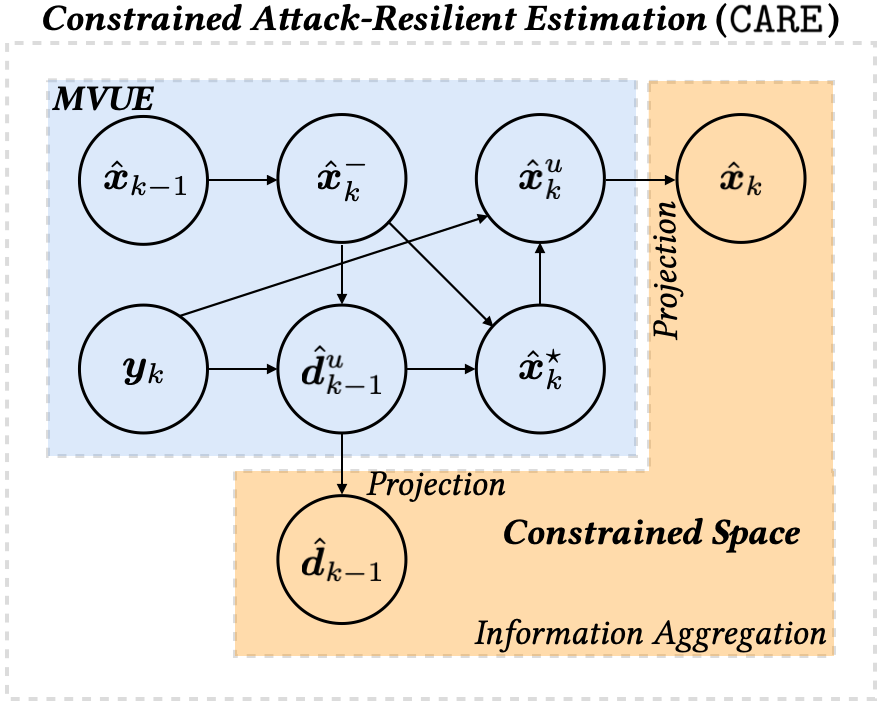}
\caption{Constrained Attack-Resilient Estimation (\texttt{CARE}). }
\label{fig: algorithm demo}
\end{center}       
\vspace{-8mm}
\end{figure}

\section{Algorithm Design}

To address the problem statement described in Section~\ref{pre}, we propose a constrained attack-resilient estimation algorithm (\texttt{CARE}), as sketched in Fig.~\ref{fig: algorithm demo}, which consists of a minimum-variance unbiased estimator (MVUE) and an information aggregation step via projection. In particular, the optimal estimation provides minimum-variance unbiased estimates, and these estimates are projected onto the constrained space eventually in the information aggregation step. We outline the essential steps of \texttt{CARE} in Section~\ref{sec:alsop} and provide a detailed derivation of the algorithm in Section~\ref{algodetail}. 

\subsection{Algorithm Statement} \label{sec:alsop}
The proposed \texttt{CARE} can be summarized as follows:
\begin{align}
    &\textbf{prediction: }
    \hat{\vx}_{k}^{-}= \mA_{k-1} \hat{\vx}_{k-1}+\mB_{k-1} \vu_{k-1} ; \label{eq: prediction} \\
    &\textbf{attack estimation: }
    \hat{\vd}_{k-1}^{u} = \mM_{k} (\vy_{k}- \mC_{k} \hat{\vx}^{-}_{k});\label{eq: d_hat_star} \\
    &\textbf{time update: }
    \hat{\vx}^{\star}_{k}=\hat{\vx}^{-}_{k}+\mG_{k-1} \hat{\vd}^{u}_{k-1} ; \label{eq: x_hat_star} \\
    &\textbf{measurement update: }
    \hat{\vx}^{u}_{k}=\hat{\vx}^{\star}_{k} +\mL_k(\vy_{k}- \mC_{k} \hat{\vx}^{\star}_{k});\label{eq: x update} \\
    &\textbf{projection update: } \nonumber
\end{align}
\vspace{-7mm}
\begin{align}
    \hat{\vd}_{k-1} =& \argmin_\vd
    (\vd-\hat{\vd}_{k-1}^{u})^\top(\mP_{k-1}^{d,u})^{-1}(\vd-\hat{\vd}_{k-1}^{u}) \nonumber\\
                    &{\rm subject\ to \ } \mathcal{A}_{k-1} \vd \leq \vb_{k-1};
    \label{eq: d_hat_sum}
    \\ \hat{\vx}_{k} =& \argmin_\vx (\vx-\hat{\vx}^{u}_{k})^\top (\mP_{k}^{x,u})^{-1}(\vx-\hat{\vx}^{u}_{k}) \nonumber \\
                    &{\rm subject\ to \ } \mathcal{B}_{k} \vx \leq \vc_{k} . \label{eq: x projection}
\end{align}


Given the previous state estimate $\hat{\vx}_{k-1}$ and its error covariance $\mP_{k-1}^x \triangleq \mathbb{E} [\tilde{\vx}_{k-1} (\tilde{\vx}_{k-1}) ^\top]$, the current state can be predicted by $\hat{\vx}_{k}^{-}$ in~\eqref{eq: prediction} under the assumption that the attack signal $\vd_{k-1}$ is absent. The unconstrained attack estimate $\hat{\vd}_{k-1}^{u}$ can be obtained by comparing the difference between the predicted output $\mC_{k}\hat{\vx}_{k}^{-}$ and the measured output $\vy_{k}$ in~\eqref{eq: d_hat_star}, where $\mM_{k}$ is the optimal filter gain that can be obtained by applying Gauss-Markov theorem, as shown in Proposition~\ref{proposition M2} later. The state prediction $\hat{\vx}_{k}^{-}$ can be updated incorporating the unconstrained attack estimate $\hat{\vd}_{k-1}^{u}$ in~\eqref{eq: x_hat_star}. The output $\vy_{k}$ is used to correct the current state estimate in~\eqref{eq: x update}, where $\mL_{k}$ is the filter gain that is obtained by minimizing the state error covariance $\mP_{k}^{x,u}$. In the information aggregation step (projection update), we apply the input constraint in~\eqref{eq: d_hat_sum} by projecting $\hat{\vd}_{k-1}^{u}$ onto the constrained space and obtain the constrained attack estimate $\hat{\vd}_{k-1}$. Similarly, the state constraint in~\eqref{eq: x projection} is applied to obtain the constrained state estimate $\hat{\vx}_{k}$. The complete algorithm is presented in Algorithm~\ref{algorithm1}.

\begin{algorithm}[ht] \small
\caption{Constrained Attack-Resilient Estimation: $[\hat{\vd}_{k-1}$, $\mP^d_{k-1}$, $\hat{\vx}_{k}$, $\mP^x_{k}]=$ \texttt{CARE}$(\hat{\vx}_{k-1},\mP^{x}_{k-1})$} \label{algorithm1}
\begin{algorithmic}[1]
\LineComment{Prediction}
\State $\hat{\vx}_{k}^{-}=\mA_{k-1} \hat{\vx}_{k-1}+\mB_{k-1} \vu_{k-1}$;
\State $\mP_{k}^{x,-}=\mA_{k-1} \mP^x_{k-1} \mA_{k-1}^\top +\mQ_{k-1}$;
\LineComment{Attack estimation}
\State $\tilde{\mR}_{k}=(\mC_{k} \mP_{k}^{x,-} \mC_{k}^\top+\mR_{k})^{-1}$;
\State $\mM_{k}=(\mG_{k-1}^\top \mC_{k}^\top \tilde{\mR}_{k} \mC_{k} \mG_{k-1})^{-1} \mG_{k-1}^\top \mC_{k}^\top \tilde{\mR}_{k}$;
\State $\hat{\vd}_{k-1}^{u}=\mM_{k} (\vy_{k}-\mC_{k} \hat{\vx}_{k}^{-})$;
\State $\mP^{d,u}_{k-1}=(\mG_{k-1}^\top \mC_{k}^\top \tilde{\mR}_{k} \mC_{k} \mG_{k-1})^{-1}$;
\State $\mP_{k-1}^{xd} =  -\mP^x_{k-1}\mA_{k-1}^\top \mC_{k}^\top \mM_{k}^\top$;
\LineComment{Time update}
\State $\hat{\vx}^\star_{k}=\hat{\vx}_{k}^{-}+\mG_{k-1} \hat{\vd}_{k-1}^{u}$;
\State $\mP^{x  \star}_{k}=\mA_{k-1}\mP_{k-1}^x \mA_{k-1}^\top +\mA_{k-1}\mP_{k-1}^{xd}\mG_{k-1}^\top$
\Statex \hspace{1.5cm}$+\mG_{k-1}(\mP_{k-1}^{xd})^\top \mA_{k-1}^\top + \mG_{k-1} \mP_{k-1}^{d,u} \mG_{k-1}^\top$
\Statex \hspace{1.5cm}$-\mG_{k-1} \mM_{k}\mC_{k}\mQ_{k-1}-\mQ_{k-1}\mC_{k}^\top \mM_{k}^\top \mG_{k-1}^\top$
\Statex \hspace{1.5cm}$+ \mQ_{k-1}$;
\State $\tilde{\mR}^\star_{k}=\mC_{k} \mP^{x  \star}_{k} \mC_{k}^\top -\mC_{k} \mG_{k-1} \mM_{k}\mR_{k} -\mR_{k} \mM_{k}^\top \mG_{k-1}^\top \mC_{k}^\top$
\Statex \hspace{1.5cm}$ +\mR_{k}$;
\LineComment{Measurement update}
\State $\mL_k=(\mP^{x  \star}_{k} \mC_{k}^\top - \mG_{k-1} \mM_{k}  \mR_{k})\tilde{\mR}^{\star \dagger}_{k}$; 
\State $\hat{\vx}^{u}_{k}=\hat{\vx}^\star_{k} +\mL_k(\vy_{k}-\mC_{k} \hat{\vx}^\star_{k})$;
\State $\mP^{x,u}_{k}= (\mI-\mL_k \mC_{k})\mG_{k-1} \mM_{k} \mR_{k}\mL_k^\top$
\Statex \hspace{1.5cm}$+ \mL_k \mR_{k} \mM_{k}^\top \mG_{k-1}^\top (\mI-\mL_k \mC_{k})^\top$
\Statex \hspace{1.5cm}$+(\mI-\mL_k \mC_{k}) \mP^{x  \star}_{k} (\mI-\mL_k \mC_{k})^\top+\mL_k \mR_{k} \mL_k^\top$;
\LineComment{Projection update}
\State $ \bm{\gamma}_{k-1}^d = \mP_{k-1}^{d,u} \bar{\mathcal{A}}_{k-1}^\top(\bar{\mathcal{A}}_{k-1} \mP_{k-1}^{d,u} \bar{\mathcal{A}}_{k-1}^\top)^{-1}$;
\State $\hat{\vd}_{k-1} = \hat{\vd}_{k-1}^{u} - \bm{\gamma}_{k-1}^d(\bar{\mathcal{A}}_{k-1}\hat{\vd}_{k-1}^{u} - \bar{\vb}_{k-1})$;
\State $\mP^{d}_{k-1}=(\mI-\bm{\gamma}_{k-1}^d\bar{\mathcal{\mA}}_{k-1})\mP^{d,u}_{k-1} (\mI-\bm{\gamma}_{k-1}^d\bar{\mathcal{A}}_{k-1}) ^\top$;
\State $ \bm{\gamma}_{k}^x = \mP_{k}^{x,u} \bar{\mathcal{B}}_{k}^\top(\bar{\mathcal{B}}_{k} \mP_{k}^{x,u} \bar{\mathcal{B}}_{k}^\top)^{-1}$;
\State $ \hat{\vx}_{k} = \hat{\vx}^u_{k} - \bm{\gamma}_{k}^x(\bar{\mathcal{B}}_{k}\hat{\vx}^u_{k} - \bar{\vc}_{k})$;
\State $\mP^{x}_{k} = (\mI-\bm{\gamma}_{k}^x \bar{\mathcal{B}}_{k})\mP^{x,u}_{k} (\mI-\bm{\gamma}_{k}^x \bar{\mathcal{B}}_{k}) ^\top$;
\end{algorithmic}
\end{algorithm}

\subsection{Algorithm Derivation} \label{algodetail}
\paragraph{Prediction}
The current state can be predicted by~\eqref{eq: prediction} under the assumption that the attack signal $\vd_{k-1} = 0$. The prediction error covariance is
\begin{align}
    \mP_{k}^{x{},-} \triangleq \mathbb{E} [\tilde{\vx}_{k}^{-}(\tilde{\vx}_{k}^{-})^\top]= \mA_{k-1} \mP^{x{}}_{k-1} \mA_{k-1}^\top +\mQ_{k-1}. \label{eq: x- err covar}
\end{align}

\paragraph{Attack estimation}
The linear attack estimator in~\eqref{eq: d_hat_star} utilizes the difference between the measured output $\vy_k$ and the predicted output $\mC_{k} \hat{\vx}^{-}_{k}$. Substituting~\eqref{eq: sys} and~\eqref{eq: prediction} into \eqref{eq: d_hat_star}, we have
\begin{align*}
     \hat{\vd}_{k-1}^{u} =  \mM_{k}
     \big( & \mC_{k}  \mA_{k-1} \tilde{\vx}_{k-1} + \mC_{k}\mG_{k-1}\vd_{k-1}
     \\ &+ \mC_{k}\vw_{k-1} + \vv_{k} \big),
\end{align*}
which is a linear function of the attack signal $\vd_{k-1}$.
Under the assumption that there is no projection update, i.e., the state and attack estimates are unconstrained, we design the optimal gain matrix $\mM_k$ such that the estimate becomes the best linear unbiased estimate (BLUE) by the following two propositions.

\begin{prop}\label{proposition M1}
Assume that there is no projection update and $\mathbb{E} [\tilde{\vx}_0]=\mathbb{E} [\tilde{\vx}^\star_0]=0$. The state estimates $\hat{\vx}_k$ and the unconstrained attack estimates $\hat{\vd}^u_k$ are unbiased for all $k$, i.e.  $\mathbb{E} [\tilde{\vx}_k]=\mathbb{E} [\tilde{\vd}^u_{k-1}]=0, \ \forall k$, if and only if $\mM_{k} \mC_{k} \mG_{k-1} = \mI$.

\noindent \textbf{Proof:}
\textit{Sufficiency:} Assuming that $\mM_{k} \mC_{k} \mG_{k-1} = \mI$, the statement can be proved by induction.
First, we will show the statement holds when $k=0$ as a base case.
By the definition, the errors of the time update and the measurement update in~\eqref{eq: x_hat_star} and~\eqref{eq: x update} are given by
\begin{align}
    \tilde{\vx}^\star_k & \triangleq  \vx_k - \hat{\vx}^\star_k = \mA_{k-1}\tilde{\vx}_{k-1} + \mG_{k-1} \tilde{\vd}^u_{k-1} + \vw_{k-1} \label{eq: xs induction proof}\\
    \tilde{\vx}^u_k & \triangleq \vx_k - \hat{\vx}^u_k = (\mI -\mL_k \mC_k)\tilde{\vx}^\star_k- \mL_k \vv_k, \label{eq: x induction proof}
\end{align}
and the error of the unconstrained attack estimate is
\begin{subequations}\label{eq: d induction proof}
\begin{align}
    \tilde{\vd}^u_{k-1} & \triangleq \vd_{k-1} - \hat{\vd}^u_{k-1} = \vd_{k-1} - \mM_{k}
     \big(  \mC_{k}  \mA_{k-1} \tilde{\vx}_{k-1} \nonumber \\
     & +\mC_{k}\mG_{k-1}\vd_{k-1} + \mC_{k}\vw_{k-1} + \vv_{k} \big)  \nonumber \\
    &=   (\mI - \mM_{k} \mC_{k} \mG_{k-1}) \vd_{k-1} \label{eq: d induction proof t1} \\
    & -  \mM_{k} \big(  \mC_{k}  \mA_{k-1} \tilde{\vx}_{k-1} + \mC_{k}\vw_{k-1} + \vv_{k} \big). \label{eq: d induction proof t2}
\end{align}
\end{subequations}
Under the assumptions that $\mathbb{E} [\tilde{\vx}_0]=\mathbb{E} [\tilde{\vx}^\star_0]=0$ and the process noise and measurement noise are zero-mean Gaussian, i.e. $\mathbb{E} [\vw_k]=\mathbb{E} [\vv_k]=0, \ \forall k$, the expectation of the term~\eqref{eq: d induction proof t2} is zero at $k=1$.
Since $\vd_{k-1}$ is deterministic for all $k$, i.e., $\mathbb{E}[{\vd_{k-1}}] \neq 0$, we have $\mathbb{E} [\tilde{\vd}^u_0]=0$ if $\mI - \mM_{1} \mC_{1} \mG_{0} = 0$, i.e., the expectation of the term~\eqref{eq: d induction proof t1} is zero at $k=1$.Then we have $\mathbb{E} [\tilde{\vx}^\star_1]=\mathbb{E} [\tilde{\vx}^u_1]=0$ by applying expectation operation on~\eqref{eq: xs induction proof} and~\eqref{eq: x induction proof}.
In the inductive step, suppose $\mathbb{E} [\tilde{\vx}^u_{k}]=\mathbb{E} [\tilde{\vx}^\star_{k}]=0$; then $\mathbb{E} [\tilde{\vd}^u_{k}]=0$ if $\mM_{k+1} \mC_{k+1} \mG_{k} = \mI$. Then, similarly, we have $\mathbb{E} [\tilde{\vx}^\star_{k+1}]=\mathbb{E} [\tilde{\vx}^u_{k+1}]=0$ by~\eqref{eq: xs induction proof} and~\eqref{eq: x induction proof}. 
Since there is no projection update, we have $\mathbb{E} [\tilde{\vx}_k] = \mathbb{E} [\tilde{\vx}^u_k]=0\  \forall k$.

\textit{Necessity:} Assuming that $\mathbb{E} [\tilde{\vx}_k]=\mathbb{E} [\tilde{\vd}_{k-1}]=0$ for all $k$, or equivalently $\mathbb{E} [\tilde{\vx}^u_k]=\mathbb{E} [\tilde{\vd}^u_{k-1}]=0$, the statement also can be proved by induction. In~\eqref{eq: d induction proof}, if $\mathbb{E} [\tilde{\vd}^u_0]=0$ for any $\vd_0$, we have $\mM_{1} \mC_{1} \mG_{0} = \mI$. Therefore, following a similar procedure, we can show that the necessity holds.
\hfill $\blacksquare$
\end{prop}

\begin{prop}\label{proposition M2}
Assume that there is no projection update and $\mathbb{E} [\tilde{\vx}_0]=\mathbb{E} [\tilde{\vx}^\star_0]=0$. The unconstrained attack estimates $\hat{\vd}^u_k$ are BLUE if 
\begin{align}
   \mM_{k} = \big(\mG_{k-1}^\top \mC_{k}^\top \tilde{\mR}_{k} \mC_{k} \mG_{k-1}\big)^{-1} \mG_{k-1}^\top  \mC_{k}^\top \tilde{\mR}_{k}, \label{proposition M}
\end{align}
where $\tilde{\mR}_{k} \triangleq (\mC_{k} \mP_{k}^{x{},-} \mC_{k}^\top + \mR_{k})^{-1}$. 

\noindent \textbf{Proof:}
Substituting~\eqref{eq: sys dyna} into~\eqref{eq: sys measure}, we have
\begin{align}
    \vy_k 
    &= \mC_k \mG_{k-1} \vd_{k-1}  \nonumber\\
    &+ \mC_k \big(\mA_{k-1}\vx_{k-1} + \mB_{k-1} \vu_{k-1} + \vw_{k-1} \big) + \vv_k . \label{eq: prop M proof 1}
\end{align}
Subtraction of $\mC_k \hat{\vx}_{k-1}$ on the both sides of~\eqref{eq: prop M proof 1} yields
\begin{align}
    \vy_k - \mC_k \hat{\vx}_{k-1} & =  \mC_k \mG_{k-1} \vd_{k-1} \nonumber \\
    & \underbrace{+ \mC_k \big(\mA_{k-1}\tilde{\vx}^-_{k-1} + \vw_{k-1} \big) + \vv_k}_{\textit{error term}}. \label{eq: prop last GMT}
\end{align}
Since the covariances of the process noise $\vw_{k-1}$ and the measurement noise $\vv_{k}$ are known, with~\eqref{eq: x- err covar}, the covariance of the error term in~\eqref{eq: prop last GMT} can be expressed as $\mC_{k} \mP_{k}^{x{},-} \mC_{k}^\top + \mR_{k}$. Applying the Gauss-Markov theorem (see Appendix~\ref{Thm:GMThm}), we can get the minimum-variance-unbiased linear estimator (BLUE) of $\vd_{k-1}$ in~\eqref{eq: d_hat_star} with $\mM_k = \big(\mG_{k-1}^\top \mC_{k}^\top \tilde{\mR}_{k} \mC_{k} \mG_{k-1}\big)^{-1} \mG_{k-1}^\top  \mC_{k}^\top \tilde{\mR}_{k}$, where $\tilde{\mR}_{k} \triangleq (\mC_{k} \mP_{k}^{x{},-} \mC_{k}^\top + \mR_{k})^{-1}$. 
\hfill $\blacksquare$
\end{prop}

\begin{rem}
The rank condition $\rank(\mC_{k} \mG_{k-1}) = n_d$ is the sufficient condition of $\mM_{k} \mC_{k} \mG_{k-1} = \mI$ needed in Proposition~\ref{proposition M1} if $\mM_{k}$ is found by~\eqref{proposition M} in Proposition~\ref{proposition M2}.
\end{rem}

The error covariance can be found by
$
    \mP_{k-1}^{d,u} = \mM_{k}\tilde{\mR}_{k}^{-1}\mM_{k}^\top=\big(\mG_{k-1}^\top \mC_{k}^\top \tilde{\mR}_{k} \mC_{k} \mG_{k-1}\big)^{-1}.
$
The cross error covariance of the state estimate and the attack estimate is
$
    \mP_{k-1}^{xd} = - \mP^x_{k-1}\mA_{k-1}^\top \mC_{k}^\top \mM_{k}^\top.
$

\paragraph{Time update}
Given the unconstrained attack estimate $\hat{\vd}^{u}_{k-1}$, the state prediction $\hat{\vx}^{-}_k$ can be updated as in~\eqref{eq: x_hat_star}. We derive the error covariance of $\hat{\vx}^{\star}_{k}$ as
\begin{align*}
    \mP_{k}^{x\star} &\triangleq \mathbb{E} \big[\tilde{\vx}^{\star}_{k}(\tilde{\vx}^{\star}_{k})^\top \big] \nonumber 
    \\&=\mA_{k-1}\mP_{k-1}^{x} \mA_{k-1}^\top +\mA_{k-1}\mP_{k-1}^{xd} \mG_{k-1}^\top \nonumber
    \\&+\mG_{k-1}\mP_{k-1}^{dx} \mA_{k-1}^\top + \mG_{k-1}\mP_{k-1}^{d,u}\mG_{k-1}^\top + \mQ_{k-1} \nonumber
    \\&-\mG_{k-1}\mM_{k}\mC_{k} \mQ_{k-1}-\mQ_{k-1}\mC_{k}^\top \mM_{k}^\top \mG_{k-1}^\top,
\end{align*}
where $\mP_{k-1}^{dx}= (\mP_{k-1}^{xd})^\top$.

\paragraph{Measurement update}
In this step, the measurement $\vy_{k}$ is used to update the propagated estimate $\hat{\vx}^\star_{k}$ as shown in~\eqref{eq: x update}.
The covariance of the state estimation error is
\begin{align*}
    \mP^{x,u}_{k} \triangleq& \mathbb{E} [(\tilde{\vx}_{k}^u)( \tilde{\vx}_{k}^u)^\top] \\
                  =& (\mI- \mL_k \mC_{k})\mG_{k-1}\mM_{k}\mR_{k}\mL_k^\top +\mL_k \mR_{k} \mL_k^\top \nonumber\\
                   &+ \mL_k \mR_{k} \mM_{k}^\top \mG_{k-1}^\top (\mI-\mL_k \mC_{k})^\top \\
                   & +(\mI-\mL_k \mC_{k}) \mP^{x \star}_{k} (\mI-\mL_k \mC_{k})^\top.
\end{align*}
The gain matrix $\mL_k$ is obtained by minimizing the trace of $\mP^{x,u}_{k}$, i.e.
$
    \min_{\mL_k} \trace (\mP^{x,u}_{k}).
$
The solution is given by
$
    \mL_k=(\mP^{x \star}_{k} \mC_{k}^\top - \mG_{k-1} \mM_{k}  \mR_{k})\tilde{\mR}^{\star \dagger}_{k},
$
where $\tilde{\mR}^\star_{k} \triangleq \mC_{k} \mP^{x \star}_{k}\mC_{k}^\top+\mR_{k}-\mC_{k}\mG_{k-1}\mM_{k}\mR_{k}-\mR_{k}\mM_{k}^\top \mG_{k-1}^\top \mC_{k}^\top$.

\paragraph{Projection update}
We are now in the position to project the estimates onto the constrained space. Apply the first constraint in~\eqref{eq: const} to the unconstrained attack estimate $\hat{\vd}_{k-1}^{u}$, and the attack estimation problem can be formulated as the following constrained convex optimization problem
\begin{align}
\begin{split}
    \hat{\vd}_{k-1} &= \argmin\limits_\vd (\vd-\hat{\vd}_{k-1}^{u})^\top \mW^d_{k-1}(\vd-\hat{\vd}_{k-1}^{u}) \\
   & \text{subject to}\  \mathcal{A}_{k-1} \vd \leq \vb_{k-1}, \label{opt ineq}
\end{split}
\end{align}
where $\mW^d_{k-1}$ can be any positive definite symmetric weighting matrix. In the current paper, we select $\mW_{k-1}^d = (\mP_{k-1}^{d,u})^{-1}$ which results in the smallest error covariance as shown in~\cite{simon2002kalman}. From Karush-Kuhn-Tucker (KKT) conditions of optimality, we can find the corresponding active constraints. We denote $\bar{\mathcal{A}}_{k}$ and $\bar{\vb}_{k}$ the rows of $\mathcal{A}_{k}$ and the elements of $\vb_k$ corresponding to the active constraints of $\mathcal{A}_{k-1} \vd \leq \vb_{k-1}$. Then~\eqref{opt ineq} becomes
\begin{align}
\begin{split}
    \hat{\vd}_{k-1} &= \argmin\limits_\vd (\vd-\hat{\vd}_{k-1}^{u})^\top (\mP_{k-1}^{d,u})^{-1}(\vd-\hat{\vd}_{k-1}^{u}) \\
   & \text{subject to}\  \bar{\mathcal{A}}_{k-1} \vd = \bar{\vb}_{k-1}. \label{eq: attack const}
\end{split}
\end{align}
The solution of~\eqref{eq: attack const} can be found by
$
    \hat{\vd}_{k-1} = \hat{\vd}_{k-1}^{u} - \bm{\gamma}_{k-1}^d(\bar{\mathcal{A}}_{k-1}\hat{\vd}_{k-1}^{u} - \bar{\vb}_{k-1}),
$
where 
\begin{align}
    \bm{\gamma}_{k-1}^d
     &\triangleq 
    \mP_{k-1}^{d,u} \bar{\mathcal{A}}_{k-1}^\top(\bar{\mathcal{A}}_{k-1}  \mP_{k-1}^{d,u} \bar{\mathcal{A}}_{k-1}^\top)^{-1}. \label{eq: gamma_d with P}
\end{align}
The attack estimation error is
\begin{align}
    \tilde{\vd}_{k-1} &= (\mI-\bm{\gamma}_{k-1}^d\bar{\mathcal{A}}_{k-1})\tilde{\vd}_{k-1}^{u} + \bm{\gamma}_{k-1}^d(\bar{\mathcal{A}}_{k-1}\vd_{k-1}- \bar{\vb}_{k-1}) \nonumber\\
    &= \hat{\vd}^u_{k-1} - \bm{\gamma}_{k-1}^d(\bar{\mathcal{A}}_{k-1}\hat{\vd}^u_{k-1} - \bar{\vb}_{k-1}).
\label{eq:dtile}
\end{align}
The error covariance can be found by
\begin{align}
    \mP^{d}_{k-1} &\triangleq {\mathbb E}[\tilde{\vd}_{k-1}\tilde{\vd}_{k-1}^\top]  \nonumber\\ &=(\mI-\bm{\gamma}_{k-1}^d\bar{\mathcal{A}}_{k-1})\mP^{d,u}_{k-1}(\mI-\bm{\gamma}_{k-1}^d\bar{\mathcal{A}}_{k-1})^\top \label{eq_pd pdu}
\end{align}
under the assumption that $\bm{\gamma}_{k-1}^d(\bar{\mathcal{A}}_{k-1}\vd_{k-1}- \bar{\vb}_{k-1}) = 0$ holds. Notice that this assumption holds when the ground truth $\vd_{k-1}$ satisfies the active constraint $\bar{\mathcal{A}}_{k-1}\vd_{k-1}= \bar{\vb}_{k-1}$.
From~\eqref{eq: gamma_d with P}, it can be verified that $\bm{\gamma}_{k-1}^d\bar{\mathcal{A}}_{k-1}\mP^{d,u}_{k-1} = \bm{\gamma}_{k-1}^d\bar{\mathcal{A}}_{k-1}\mP^{d,u}_{k-1}(\bm{\gamma}_{k-1}^d\bar{\mathcal{A}}_{k-1})^\top$. Therefore, from~\eqref{eq_pd pdu} we have
\begin{align}
    \mP^{d}_{k-1}
    =& \mP^{d,u}_{k-1} - \bm{\gamma}_{k-1}^d\bar{\mathcal{A}}_{k-1} \mP^{d,u}_{k-1} -   \mP^{d,u}_{k-1} (\bm{\gamma}_{k-1}^d\bar{\mathcal{A}}_{k-1})^\top \nonumber \\
    & +  \bm{\gamma}_{k-1}^d\bar{\mathcal{A}}_{k-1}\mP^{d,u}_{k-1}(\bm{\gamma}_{k-1}^d\bar{\mathcal{A}}_{k-1})^\top \nonumber\\
    =&(\mI-\bm{\gamma}_{k-1}^d\bar{\mathcal{A}}_{k-1})\mP^{d,u}_{k-1}. \label{eq_pd pdu short}
\end{align}
Similarly, applying the second constraint in~\eqref{eq: const} to the unconstrained state estimate $\hat{\vx}_{k}^u$, we formalize the state estimation problem as follows:
\begin{align}
\begin{split}
    \hat{\vx}_{k} &= \argmin_{\vx} (\vx-\hat{\vx}^u_{k})^\top \mW^x_k(\vx-\hat{\vx}^u_{k}) \\
   & \text{subject to}\  \mathcal{B}_{k} \vx \leq \vc_{k},
\end{split}
\label{e035.0}
\end{align}
where we select $\mW^x_{k} = (\mP^{x,u}_{k})^{-1}$ for the smallest error covariance. We denote $\bar{\mathcal{B}}_{k}$ and $\bar{\vc}_{k}$ the rows of $\mathcal{B}_{k}$ and the elements of $\vc_k$ corresponding to the active constraints of $\mathcal{B}_{k} \vx \leq \vc_{k}$. Using the active constraints, we reformulate~\eqref{e035.0} as follows:
\begin{align}
\begin{split}
    \hat{\vx}_{k} &= \argmin\limits_\vx (\vx-\hat{\vx}^u_{k})^\top (\mP^{x,u}_{k})^{-1}(\vx-\hat{\vx}^u_{k}) \\
   & \text{subject to}\  \bar{\mathcal{B}}_{k} \vx = \bar{\vc}_{k}.
\end{split}\label{e035}
\end{align}
The solution of~\eqref{e035} is given by
$
    \hat{\vx}_{k} = \hat{\vx}^u_{k} - \bm{\gamma}_{k}^x(\bar{\mathcal{B}}_{k}\hat{\vx}^u_{k} - \bar{\vc}_{k}),
$
where 
\begin{align} \label{eq: gammax def}
    \bm{\gamma}_{k}^x
    &\triangleq
    \mP^{x,u}_{k} \bar{\mathcal{B}}_{k}^\top(\bar{\mathcal{B}}_{k} \mP^{x,u}_{k}  \bar{\mathcal{B}}_{k}^\top)^{-1}.
\end{align}
Under the assumption that $\bm{\gamma}_{k}^x(\bar{\mathcal{B}}_{k}\hat{\vx}^u_{k} - \bar{\vc}_{k})=0$ holds, the state estimation error covariance can be expressed as
\begin{align}
    \mP^{x}_{k} &=\bar{ \bm{\Gamma}}_k \mP^{x,u}_{k} \bar{ \bm{\Gamma}}_k ^\top, \label{eq:Px projected}
\end{align}
where $\bar{ \bm{\Gamma}}_k \triangleq \mI-\bm{\gamma}_{k}^x\bar{\mathcal{B}}_{k}$.
Notice that this assumption holds when the ground truth $\vx_k$ satisfies the active constraint $\bar{\mathcal{B}}_{k}\vx_k= \bar{\vc}_{k}$.

\section{Performance and Stability Analysis}\label{analysis} 

In Section~\ref{sec:proe}, we show that the projection induced by inequality constraints improves attack-resilient estimation accuracy and detection performance by decreasing estimation errors and the false negative rate in attack detection. Notice that the estimate $ \hat{\vd}_{k-1}$ and the ground truth $\vd_{k-1}$ satisfy the active constraint $\bar{\mathcal{A}}_{k-1}\hat{\vd}_{k-1}- \bar{\vb}_{k-1}=0$ in~\eqref{eq: attack const} and the inequality constraint $\mathcal{A}_{k-1}\vd_{k-1} \leq \vb_{k-1}$ in~\eqref{eq: const}, respectively. However, it is uncertain whether the ground truth satisfies the active constraints or not. In this case, from~\eqref{eq:dtile} we have
\begin{align}
    \mathbb{E}[\tilde{\vd}_{k-1}] &= \bm{\gamma}_{k-1}^d(\bar{\mathcal{A}}_{k-1}\vd_{k-1}- \bar{\vb}_{k-1}) \neq 0  \label{eq:biased d}.
\end{align}
A similar statement holds for the state estimation error:
\begin{align}
    \mathbb{E}[\tilde{\vx}_{k}] &= \bm{\gamma}_{k}^x(\bar{\mathcal{B}}_{k} {\vx}_{k} - \bar{\vc}_{k}) \neq 0.\label{eq:biased x}
\end{align}
These considerations indicate that the projection potentially induces biased estimates, rendering the traditional stability analysis for unbiased estimation invalid. In this context, we will prove that estimation errors of the \texttt{CARE} are practically exponentially stable in mean square which will be proven in Section~\ref{sec:stab}.

\subsection{Performance Analysis}\label{sec:proe}
For the analysis of the performance through the projection, we first decompose the state estimation error $\tilde{\vx}_k$ into two orthogonal spaces as follows:
\begin{align} \label{decompose x tilde}
    \tilde{\vx}_k &= (\mI-\bm{\gamma}_k^x \bar{\mathcal{B}}_k) \tilde{\vx}_k + \bm{\gamma}_k^x \bar{\mathcal{B}}_k \tilde{\vx}_k.
\end{align}
We will show that the errors in the space $\mI-\bm{\gamma}_k^x\bar{\mathcal{B}}_k$ remain identical after the projection, while the errors in the space $\bm{\gamma}_k^x\bar{\mathcal{B}}_k$ reduce through the projection, as in Lemma~\ref{lem22}.

\begin{lem} \label{lem22}
The decomposition of $\tilde{\vx}_{k}$ in the space $\mI-\bm{\gamma}_k^x \bar{\mathcal{B}}_k$ is equal to that of $\tilde{\vx}_{k}^u$, and the decomposition of $\tilde{\vx}_{k}$ in the space $\bm{\gamma}_k^x\bar{\mathcal{B}}_k$ is equal to that of $\tilde{\vx}_{k}^u$ scaled by $\bm{\alpha}_k$, i.e.
\begin{align}
    (\mI-\bm{\gamma}_k^x \bar{\mathcal{B}}_k)\tilde{\vx}_{k} &=(\mI-\bm{\gamma}_k^x \bar{\mathcal{B}}_k)\tilde{\vx}_{k}^u \label{eq: lem_statement 1}\\
        \bm{\gamma}_k^x\bar{\mathcal{B}}_k\tilde{\vx}_{k} &= \bm{\alpha}_k\bm{\gamma}_k^x\bar{\mathcal{B}}_k\tilde{\vx}_{k}^u,\label{eq: lem_statement 2}
    \end{align}
where $\bm{\alpha}_k=\diag{(\bm{\alpha}_k^1,\cdots,\bm{\alpha}_k^n)}$, and 
    $$\bm{\alpha}_k^i \triangleq (\bm{\gamma}_k^x\bar{\mathcal{B}}_k\tilde{\vx}_k)(i)((\bm{\gamma}_k^x\bar{\mathcal{B}}_k\tilde{\vx}_k^u)(i))^\dagger \in [0,1)$$
for $i=1,\cdots,n$.
Similarly, it holds that
$(\mI-\bm{\gamma}_k^d \bar{\mathcal{A}}_k)\tilde{\vd}_{k} =(\mI-\bm{\gamma}_k^d \bar{\mathcal{A}}_k)\tilde{\vd}_{k}^u$ and 
$\bm{\gamma}_k^d\bar{\mathcal{A}}_k\tilde{\vd}_{k} = \bm{\kappa}_k\bm{\gamma}_k^d\bar{\mathcal{A}}_k\tilde{\vd}_{k}^u$,
    where $\bm{\kappa}_k=\diag{(\bm{\kappa}_k^1,\cdots,\bm{\kappa}_k^n)}$, and 
    $
    \bm{\kappa}_k^i \triangleq (\bm{\gamma}_k^d\bar{\mathcal{A}}_k\tilde{\vd}_k)(i)((\bm{\gamma}_k^d\bar{\mathcal{A}}_k\tilde{\vd}_k^u)(i))^\dagger \in [0,1) 
    $
for $i=1,\cdots,n$.

\noindent \textbf{Proof:}
    The relationship in~\eqref{eq: lem_statement 1} can be obtained by applying $\bar{\mathcal{B}}_k\hat{\vx}_{k}=\bar{\vc}_k$ to
\begin{align*}
    \tilde{\vx}_{k} &= \vx_k - \hat{\vx}_{k} = \vx_k - (\hat{\vx}_{k}^u-\bm{\gamma}_k^x(\bar{\mathcal{B}}_k\hat{\vx}_{k}^u-\bar{\vc}_k))\\
    &=\tilde{\vx}_{k}^u+\bm{\gamma}_k^x(\bar{\mathcal{B}}_k\hat{\vx}_{k}^u-\bar{\vc}_k)\\
    &=\tilde{\vx}_{k}^u+\bm{\gamma}_k^x(\bar{\mathcal{B}}_k\hat{\vx}_{k}^u-\bar{\mathcal{B}}_k\hat{\vx}_{k})\\
    &=\tilde{\vx}_{k}^u-\bm{\gamma}_k^x(\bar{\mathcal{B}}_k\tilde{\vx}_{k}^u-\bar{\mathcal{B}}_k\tilde{\vx}_{k}),
\end{align*}
which implies~\eqref{eq: lem_statement 1}. The solution of $\bar{\mathcal{B}}_k \vx \leq \bar{\vc}_k$ defines a closed convex set ${\mathcal C}_k$. The point $\hat{\vx}_{k}^u$ is not an element of the convex set. The point $\hat{\vx}_{k}$ has the minimum distance from $\hat{\vx}_{k}^u$ with metric $d(a,b) \triangleq \|a-b\|_{\mW_k^x}$ in the convex set ${\mathcal C}_k$ by~\eqref{e035}. Since the solution $\hat{\vx}_k$ is in the closed set ${\mathcal C}_k$, and $\bm{\gamma}_k^x\bar{\mathcal{B}}_k$ is a weighted projection with weight $\mW_k^x$, the relationship~\eqref{eq: lem_statement 2} holds. The statements for attack estimation errors can be proven by a similar procedure, which is omitted here. 
\hfill $\blacksquare$
\end{lem}

With the results from Lemma~\ref{lem22}, we can show that the projection reduces the estimation errors and the error covariances, as formulated in Theorem~\ref{the1}.

\begin{thm}
\texttt{CARE} reduces the state and attack estimation errors and their error covariances from the unconstrained algorithm, i.e., $\|\tilde{\vx}_{k}\|\leq \|\tilde{\vx}_{k}^u\|$ and $\|\tilde{\vd}_{k}\|\leq \|\tilde{\vd}_{k}^u\|$, $\mP_k^x \leq \mP_k^{x,u}$ and $\mP_k^d \leq \mP_k^{d,u}$. Strict inequality holds if $\rank(\bar{\mathcal{B}}_k) \neq 0$, and $\rank(\bar{\mathcal{A}}_k) \neq 0$, respectively.
\label{the1}

\noindent \textbf{Proof:}
The statement for $\|\tilde{\vx}_{k}\|\leq \|\tilde{\vx}_{k}^u\|$  is the direct result of Lemma~\ref{lem22}, where strict inequality holds if $\bm{\alpha}_k^i \neq 0$ for some $i$. The statement for $\|\tilde{\vd}_{k}\|\leq \|\tilde{\vd}_{k}^u\|$ can be proved by a similar procedure.
To show the rest of the properties, we first identify the equality
\begin{align}
    (\mI-\bm{\gamma}_k^x \bar{\mathcal{B}}_k)^\top\bm{\gamma}_k^x \bar{\mathcal{B}}_k=0. \label{cle2}
\end{align}
Since we have $\bar{\mathcal{B}}_k\bm{\gamma}_k^x= \mI$ by~\eqref{eq: gammax def}, it holds that 
$\bm{\gamma}_k^x\bar{\mathcal{B}}_k\bm{\gamma}_k^x = \bm{\gamma}_k^x$, and $\bar{\mathcal{B}}_k\bm{\gamma}_k^x\bar{\mathcal{B}}_k = \bar{\mathcal{B}}_k$, i.e. $\bm{\gamma}_k^x = \bar{\mathcal{B}}_k^\dagger$.
Then, we have $\bar{\mathcal{B}}_k^\top (\bm{\gamma}_k^x)^\top \bm{\gamma}_k^x=\bm{\gamma}_k^x$, which implies
$\tilde{\vx}_{k}^\top(\mI-\bm{\gamma}_k^x \bar{\mathcal{B}}_k)^\top\bm{\gamma}_k^x \bar{\mathcal{B}}_k\tilde{\vx}_{k}=\tilde{\vx}_{k}^\top(\bm{\gamma}_k^x\bar{\mathcal{B}}_k-\bar{\mathcal{B}}_k^\top (\bm{\gamma}_k^x)^\top \bm{\gamma}_k^x\bar{\mathcal{B}}_k)\hat{\vx}_{k}=0$. Notice that~\eqref{cle2} holds for $(\tilde{\vx}_{k}^u)^\top(\mI-\bm{\gamma}_k^x \bar{\mathcal{B}}_k)^\top\bm{\gamma}_k^x \bar{\mathcal{B}}_k\tilde{\vx}_{k}^u=0$ as well.
Similar to~\eqref{eq_pd pdu short}, we have
$\mP_{k}^{x} = (\mI-\bm{\gamma}_k^x \bar{\mathcal{B}}_k)\mP_k^{x,u} = \mP_k^{x,u}- \bm{\gamma}_k^x \bar{\mathcal{B}}_k\mP_k^{x,u}$. Given that $\bm{\gamma}_k^x \bar{\mathcal{B}}_k\mP_k^{x,u}=\mP_k^{x,u} \bar{\mathcal{B}}_{k}^\top(\bar{\mathcal{B}}_{k} \mP_k^{x,u} \bar{\mathcal{B}}_{k}^\top)^{-1}\bar{\mathcal{B}}_k \mP_k^{x,u}>0$ is positive definite, we have the desired result $\mP_{k}^{x} < \mP_k^{x,u}$.
The relation for $\mP_k^d$ can be obtained by a similar procedure.
\hfill $\blacksquare$
\end{thm}

The properties in Theorem~\ref{the1} are desired for accurate estimation as well as attack detection. More specifically, since the false negative rate of a $\chi^2$ attack detector is a function of the estimate $\hat{\bm{\sigma}}_k$ and the covariance $\bm{\Sigma}_k$ as in~\eqref{eq: Fneg define}, more accurate estimations can reduce the false negative rate under the following assumption.

\begin{assum}
\label{assum: d>chi}
In the presence of the attack ($\vd_k \neq 0$), the following two conditions hold:
(i) $\|\tilde{\vd}^u_k\|< \frac{1}{2}\|\vd_k\|$, and (ii) the ground truth $\vd_k$ satisfies the condition $\vd_k^\top (\mP_{k}^{d,u})^{-1}\vd_k > \chi^2_{df}(\alpha)$.
\end{assum}

\begin{rem}
Assumption~\ref{assum: d>chi} implies that the unconstrained attack estimation error $\tilde{\vd}^u_k$ is small with respect to the ground truth $\vd_k$, and the normalized ground truth attack signal is larger than $\chi^2_{df}(\alpha)$; otherwise, it cannot be distinguished from the noise.
Notice that Assumption~\ref{assum: d>chi} is only considered for smaller false negative rates (Theorem~\ref{thm_Fneg}), but not for the estimation performance (Theorem~\ref{the1}) and stability analysis in Section~\ref{sec:stab}, where we will show the stability of the attack estimation error $\tilde{\vd}_k$ (Theorem~\ref{the2}) which renders the stability of $\tilde{\vd}^u_k$.
\end{rem}

According to~\eqref{eq: Fneg define}, we denote the false negative rates of the proposed \texttt{CARE} and the unconstrained algorithm as $F_{neg}(\{ \hat{\vd}_k \}, \{\mP_{k}^{d} \})$ and $F_{neg}(\{ \hat{\vd}_k^u\}, \{ \mP_{k}^{d,u}\})$, respectively. The following Theorem~\ref{thm_Fneg} demonstrates that the false negative rate of \texttt{CARE} is less or equal to that of the unconstrained algorithm. 

\begin{thm} \label{thm_Fneg}
Under Assumption~\ref{assum: d>chi}, given a set of attack vectors $\{ \vd_k\}$, the following inequality holds
\begin{align}
    F_{neg}(\{\hat{\vd}_k\}, \{\mP_{k}^{d}\}) &\leq F_{neg}(\{\hat{\vd}_k^u\}, \{\mP_{k}^{d,u}\}). \label{eq: small_neg}
\end{align}
\textbf{Proof: }
To prove~\eqref{eq: small_neg} is equivalent to showing that the number of false negative test results of \texttt{CARE} is less or equal to that of the unconstrained algorithm
\begin{align}
   \sum_{k} (\bm{1}_k) \leq \sum_{k} (\bm{1}_k^u). \label{eq: 2ineqs}
\end{align}
If there is no projection ($\bm{\gamma}_{k}^d = 0$), it holds that $\hat{\vd}_k = \hat{\vd}_k^u$ and $\mP_{k}^{d} = \mP_{k}^{d,u}$.
And, if there is no attack ($\vd_k = 0$), it holds that $\bm{1}_k=\bm{1}_k^u=0$. Therefore, we have 
\begin{align}
     \sum_{k\in \mathcal{K}_0} (\bm{1}_k) = \sum_{k\in \mathcal{K}_0} (\bm{1}_k^u), \label{eq: 2ineqs_P0}
\end{align}
where $\mathcal{K}_0 \triangleq \{ k \mid \bm{\gamma}_{k}^d = 0  \textit{ or } \vd_k = 0\}$.
In the rest of the proof, we consider the case for $k \in \mathcal{K} \triangleq \{ k \mid \bm{\gamma}_{k}^d \neq 0 \textit{ and } \vd_k \neq 0\}$.
Rewriting the normalized test value from \texttt{CARE} by substituting $\mP_{k}^{d}$ with $(\mI-\bm{\gamma}_k^d \bar{\mathcal{A}}_k)\mP_{k}^{d,u}  (\mI-\bm{\gamma}_k^d \bar{\mathcal{A}}_k)^\top$ according to~\eqref{eq_pd pdu}, we have the following:
\begin{align}
\hat{\vd}_k^\top &(\mP_{k}^{d})^{-1} \hat{\vd}_k
 = \hat{\vd}_k^\top \big( (\mI-\bm{\gamma}_k^d \bar{\mathcal{A}}_k)\mP_{k}^{d,u}  (\mI-\bm{\gamma}_k^d \bar{\mathcal{A}}_k)^\top \big)^{-1} \hat{\vd}_k \nonumber\\
 =& \big( (\mI-\bm{\gamma}_k^d \bar{\mathcal{A}}_k)^{-1}\hat{\vd}_k\big)^\top(\mP_{k}^{d,u} )^{-1} \big( (\mI-\bm{\gamma}_k^d \bar{\mathcal{A}}_k)^{-1}\hat{\vd}_k\big) \nonumber\\
 =& \big( \hat{\vd}_k^u + (\mI-\bm{\gamma}_k^d \bar{\mathcal{A}}_k)^{-1} \bm{\gamma}_{k}^d \bar{\vb}_{k} \big)^\top(\mP_{k}^{d,u} )^{-1} \nonumber\\
& \times \big(   \hat{\vd}_k^u + (\mI-\bm{\gamma}_k^d \bar{\mathcal{A}}_k)^{-1} \bm{\gamma}_{k}^d \bar{\vb}_{k}\big), \label{eq_norm care test value}
\end{align}
where $(\mI-\bm{\gamma}_k^d \bar{\mathcal{A}}_k)^{-1}\hat{\vd}_k = \hat{\vd}_k^u + (\mI-\bm{\gamma}_k^d \bar{\mathcal{A}}_k)^{-1} \bm{\gamma}_{k}^d \bar{\vb}_{k}$ has been applied. Now we expand and rearrange~\eqref{eq_norm care test value}, resulting in the following:
\begin{align}
&\hat{\vd}_k^\top (\mP_{k}^{d})^{-1} \hat{\vd}_k = 
( \hat{\vd}_k^u)^\top(\mP_{k}^{d,u} )^{-1} \hat{\vd}_k^u  \nonumber\\ 
+&  \big((\mI-\bm{\gamma}_k^d \bar{\mathcal{A}}_k)^{-1} \bm{\gamma}_{k}^d \bar{\vb}_{k}  \big)^\top(\mP_{k}^{d,u} )^{-1}   \big((\mI-\bm{\gamma}_k^d \bar{\mathcal{A}}_k)^{-1} \bm{\gamma}_{k}^d \bar{\vb}_{k}  \big) \nonumber\\
+& 2( \hat{\vd}_k^u)^\top(\mP_{k}^{d,u} )^{-1} ((\mI-\bm{\gamma}_k^d \bar{\mathcal{A}}_k)^{-1} \bm{\gamma}_{k}^d \bar{\vb}_{k})\nonumber \\
=& ( \hat{\vd}_k^u)^\top(\mP_{k}^{d,u} )^{-1}    \hat{\vd}_k^u \nonumber\\
+&  \big(\bm{\gamma}_{k}^d \bar{\vb}_{k}  \big)^\top\big( (\mI-\bm{\gamma}_k^d \bar{\mathcal{A}}_k)\mP_{k}^{d,u}  (\mI-\bm{\gamma}_k^d \bar{\mathcal{A}}_k)^\top \big)^{-1}  \bm{\gamma}_{k}^d \bar{\vb}_{k} \nonumber\\
+& 2( \hat{\vd}_k^u)^\top  \big( (\mI-\bm{\gamma}_k^d \bar{\mathcal{A}}_k)\mP_{k}^{d,u} \big)^{-1}  \bm{\gamma}_{k}^d \bar{\vb}_{k}. \label{eq:PFprof +step expand}
\end{align}
Applying~\eqref{eq_pd pdu} and \eqref{eq_pd pdu short} to~\eqref{eq:PFprof +step expand}, we have
\begin{align}
&\hat{\vd}_k^\top (\mP_{k}^{d})^{-1} \hat{\vd}_k = 
  (\hat{\vd}_k^u)^\top(\mP_{k}^{d,u} )^{-1} \hat{\vd}_k^u \nonumber\\
+& \underbrace{\big(\bm{\gamma}_{k}^d \bar{\vb}_{k}  \big)^\top ( \mP_{k}^{d})^{-1} \bm{\gamma}_{k}^d \bar{\vb}_{k}+ 2( \hat{\vd}_k^u)^\top  (\mP_{k}^{d})^{-1}  \bm{\gamma}_{k}^d \bar{\vb}_{k}}_{\triangleq \ residue \ (res.)}. \label{eq:PFprof +res.}
\end{align}
Since $\hat{\vd}_k$ satisfies the input active constraint, we can substitute $\bar{\vb}_{k}$ with $\bar{\mathcal{A}}_k\hat{\vd}_k$. Then the residue defined in~\eqref{eq:PFprof +res.} can be written as follows:
\begin{align}
&res. =\big(\bm{\gamma}_{k}^d \bar{\mathcal{A}}_k\hat{\vd}_k  \big)^\top ( \mP_{k}^{d})^{-1} \bm{\gamma}_{k}^d \bar{\mathcal{A}}_k\hat{\vd}_k \nonumber\\
& \ \ \ \ \ \ \ \ + 2( \hat{\vd}_k^u)^\top  (\mP_{k}^{d})^{-1}  \bm{\gamma}_{k}^d \bar{\mathcal{A}}_k\hat{\vd}_k. \label{eq: res}
\end{align}
Expanding and rearranging~\eqref{eq: res}, we have the following:
\begin{align}
&res. = 2 \vd_k^\top \mP'_k \vd_k
- 2 \tilde{\vd}_k^\top    \mP'_k  {\vd}_k 
- 2 (\tilde{\vd}_k^u)^\top  \mP'_k   {\vd}_k \label{eq:res 1}\\
&+  2 \tilde{\vd}_k^\top \mP'_k    \tilde{\vd}_k
+ \|\bm{\gamma}_{k}^d \bar{\mathcal{A}}_k\|^2  \tilde{\vd}_k^\top    ( \mP_{k}^{d})^{-1}     \tilde{\vd}_k  \label{eq:res 2}\\
&+ \|\bm{\gamma}_{k}^d \bar{\mathcal{A}}_k\|^2 \vd_k^\top    ( \mP_{k}^{d})^{-1}     \vd_k
- 2 \|\bm{\gamma}_{k}^d \bar{\mathcal{A}}_k\|^2  {\vd}_k^\top    ( \mP_{k}^{d})^{-1}     \tilde{\vd}_k,  \label{eq:res 3}
\end{align}
where $\mP'_k \triangleq (\bm{\gamma}_{k}^d \bar{\mathcal{A}}_k)^\top( \mP_{k}^{d})^{-1} > 0$.
Using the result $\|\tilde{\vd}_k\|< \|\tilde{\vd}^u_k\|$ from Theorem~\ref{the1} and the first inequality in Assumption~\ref{assum: d>chi}, we obtain $\|\tilde{\vd}\|< \|\tilde{\vd^u}\|< \frac{1}{2}\|\vd\|$.
Then we have $res. > 0$, since~\eqref{eq:res 1} to~\eqref{eq:res 3} are positive, respectively.
Therefore, from~\eqref{eq:PFprof +res.}, we have
\begin{align} \label{eq:PF_pre_result}
( \hat{\vd}_k^u)^\top(\mP_{k}^{d,u} )^{-1}    \hat{\vd}_k^u <
\hat{\vd}_k^\top (\mP_{k}^{d})^{-1} \hat{\vd}_k.
\end{align}
Considering the condition in~\eqref{eq:PF_pre_result}, we can divide the set $\mathcal{K} = \cup_{i=1}^3 \mathcal{K}_i$ into three partitions as follows:
\begin{align*}
\mathcal{K}_1 &\triangleq \big\{ k \mid (\hat{\vd}_k^u)^\top(\mP_{k}^{d,u} )^{-1}  \hat{\vd}_k^u <
\hat{\vd}_k^\top (\mP_{k}^{d})^{-1} \hat{\vd}_k \leq \chi^2_{df}(\alpha)  \big\}\\
\mathcal{K}_2 &\triangleq  \big\{ k \mid \chi^2_{df}(\alpha) < ( \hat{\vd}_k^u)^\top(\mP_{k}^{d,u} )^{-1}    \hat{\vd}_k^u <
\hat{\vd}_k^\top (\mP_{k}^{d})^{-1} \hat{\vd}_k  \big\} \\
\mathcal{K}_3 &\triangleq  \big\{ k \mid ( \hat{\vd}_k^u)^\top(\mP_{k}^{d,u} )^{-1}    \hat{\vd}_k^u
\leq \chi^2_{df}(\alpha)<
\hat{\vd}_k^\top (\mP_{k}^{d})^{-1} \hat{\vd}_k  \big\}.
\end{align*}
According to~\eqref{eq: indicator}, we have 
\begin{align}
     \sum_{k\in \mathcal{K}_i} (\bm{1}_k) &= \sum_{k\in \mathcal{K}_i} (\bm{1}_k^u) \textit{ for } i = 1,2 \textit{ and  }  \label{eq: 2ineqs_P12} \\ 
     \sum_{k\in \mathcal{K}_3} (\bm{1}_k) &< \sum_{k\in \mathcal{K}_3} (\bm{1}_k^u). \label{eq: 2ineqs_P3}
\end{align}
Therefore, from~\eqref{eq: 2ineqs_P0}, \eqref{eq: 2ineqs_P12} and~\eqref{eq: 2ineqs_P3} we conclude that~\eqref{eq: 2ineqs} holds, which completes the proof.
\hfill $\blacksquare$
\end{thm}

\subsection{Stability Analysis}\label{sec:stab}
Although the projection reduces the estimation errors and their error covariances as shown in Theorem~\ref{the1}, it trades the unbiased estimation off according to~\eqref{eq:biased d} and~\eqref{eq:biased x}. In the absence of the projection, Algorithm~\ref{algorithm1} reduces to the algorithm in~\cite{yong2016unified}, which is an unbiased estimation, while the traditional stability analysis for unbiased estimation becomes invalid after the projection is applied.

To prove the recursive stability of the biased estimation, it is essential to construct a recursive relation between the current estimation error $\tilde{\vx}_{k}$ and the previous estimation error $\tilde{\vx}_{k-1}$. However, the construction is not straightforward compared to that in filtering with equality constraints~\cite{simon2002kalman,yong2015simultaneous} or filtering without constraints~\cite{anderson1981detectability,yong2016unified}. Especially, it is difficult to find the exact recursive relation between $\tilde{\vx}_{k}$ and $\tilde{\vx}_{k}^u$, since $\tilde{\vx}_{k}$ is also a function of $\hat{\vx}^u_{k}$, i.e. $\tilde{\vx}_{k} = \tilde{\vx}^u_{k} - \bm{\gamma}_{k}^x(\bar{\mathcal{B}}_{k}\hat{\vx}^u_{k} - \bar{\vc}_{k})$. Then, we have $\tilde{\vx}_{k} \neq (\mI-\bm{\gamma}_{k}^x\bar{\mathcal{B}}_{k})\tilde{\vx}^u_{k}$, since the inequality $\bar{\mathcal{B}}_k \vx_k \leq \bar{\vc}_k$ holds.
To address this issue, we decompose the estimation error $\tilde{\vx}_{k}$ into two orthogonal spaces as in~\eqref{decompose x tilde}. By Lemma~\ref{lem22}, \eqref{decompose x tilde} becomes
$
	\tilde{\vx}_{k} =  \bm{\Gamma}_{k}\tilde{\vx}_{k}^u,  
$
where $ \bm{\Gamma}_{k} \triangleq (\mI-\bm{\gamma}_k^x \bar{\mathcal{B}}_k)+\bm{\alpha}_k\bm{\gamma}_k^x \bar{\mathcal{B}}_k$. Note that $\bm{\alpha}_k$ is an unknown matrix and thus cannot be used for the algorithm. We use it only for analytical purposes. Now under the following assumptions, we present the stability analysis of the proposed Algorithm~\ref{algorithm1}.
\begin{assum}\label{assum_bounded}
   We have $\rank(\mathcal{B}_k) < n$ $\forall k$.
    There exist $\bar{a}$, $\bar{c}_y$, $\bar{g}$, $\bar{m}$, $\ubar{q}$, $\ubar{\beta}$, $\bar{\beta}$ $>0$, such that the following holds for all $k \geq 0$:
    \begin{align*}
        &\|\mA_k\| \leq \bar{a},
        &&\|\mC_{k}\| \leq \bar{c}_y,
        &&\|\mG_{k}\| \leq \bar{g},\\
        & \|\mM_{k}\| \leq \bar{m},
        &&\mQ_k\geq \ubar{q} \mI.
\end{align*}
\end{assum}

\begin{rem}
In Assumption~\ref{assum_bounded}, it is assumed that $\rank(\mathcal{B}_k) < n$ $\forall k$, i.e., the number of the state constraints are less than the number of state variables. The rest of Assumption~\ref{assum_bounded} is widely used in the literature on extended Kalman filtering~\cite{kluge2010stochastic} and nonlinear input and state estimation~\cite{kim2017attack}.
\end{rem}

To show the boundedness of the unconstrained state error covariance $\mP_k^{x,u}$, we first define the matrices
$\bar{\mA}_{k-1}  \triangleq (\mI-\mG_{k-1}\mM_{k}\mC_{k})\mA_{k-1}$ and $\tilde{\mA}_{k-1} \triangleq (\mI-\mG_{k-1}\mM_{k}(\mC_{k}\mG_{k-1}\mM_{k})^{-1}\mC_{k})\bar{\mA}_{k-1}\bar{ \bm{\Gamma}}_{k-1}$.
\begin{thm}\label{the3}
Let the pair $(\mC_{k}, \tilde{\mA}_{k-1})$ be uniformly detectable\footnote{Please refer to~\cite{anderson1981detectability} for the definition of uniform detectability.}, then the unconstrained state error covariance $\mP_k^{x,u}$ is bounded, i.e., there exist non-negative constants $\ubar{p}$ and $\bar{p}$ such that $\ubar{p} \mI \leq \mP_k^{x,u} \leq \bar{p} \mI$ for all $k$.

\noindent  \textbf{Proof: }   
The unconstrained state estimation error can be found by
\begin{align}
    \tilde{\vx}^u_{k} =& (\mI-\mL_k \mC_{k}) \bar{\mA}_{k-1}\tilde{\vx}_{k-1} \nonumber \\
    &+(\mI-\mL_k\mC_{k})\bar{\vw}_{k-1}+\bar{\mL}_k\vv_{k},
 \label{eq:new unc_state error}
\end{align}
where $\bar{\vw}_{k-1} \triangleq (\mI- \mG_{k-1}\mM_{k}\mC_{k})\vw_{k-1}$, and $\bar{\mL}_k \triangleq \mL_k \mC_{k} \mG_{k-1} \mM_{k}-\mL_k-\mG_{k-1}\mM_{k}$. Therefore, the update law of unconstrained covariance is calculated from~\eqref{eq:new unc_state error} and \eqref{eq:Px projected} as follows:
    \begin{align}
        \mP_{k}^{x,u} =& (\mI-\mL_k\mC_{k})\bar{\mA}_{k-1}\bar{ \bm{\Gamma}}_{k-1} \mP_{k-1}^{x,u} \bar{ \bm{\Gamma}}_{k-1}^\top \nonumber\\
        &\times \bar{\mA}_{k-1}^ \top(\mI-\mL_k\mC_{k})^\top 
        +\bar{\mL}_k\mR_{k}\bar{\mL}_k^\top \nonumber\\ &+(\mI-\mL_k\mC_{k})\bar{\mQ}_{k-1}(\mI-\mL_k\mC_{k})^\top,
        \label{eq: new P (Gamma)}
    \end{align}
where $\bar{\mQ}_{k-1} \triangleq  \mathbb{E}[\bar{\vw}_{k-1} (\bar{\vw}_{k-1})^\top] $.
The covariance update law~\eqref{eq: new P (Gamma)} is identical to the covariance update law of the Kalman filtering solution of the transformed system
\begin{align}
\begin{split}
    \vx_{k} &= \bar{\mA}_{k-1}\bar{ \bm{\Gamma}}_{k-1}\vx_{k-1}+\hat{\vw}_{k-1}\\
    \vy_{k} &= \mC_{k}\vx_{k}+\vv_{k},\label{eq:equiv1}
\end{split}
\end{align}
where $\hat{\vw}_{k-1} \triangleq - \mG_{k-1}\mM_{k}\mC_{k}\vw_{k-1} - \mG_{k-1}\mM_{k}\vv_{k}+\vw_{k-1}$. However, in the transformed system, the process noise and measurement noise are correlated, i.e., $\mathbb{E}[\hat{\vw}_{k-1}\vv_{k}^\top]=-\mG_{k-1}\mM_{k}\mR_{k}\neq 0$. To decouple the noises, we add a zero term $\mZ_k(\vy_{k}-\mC_{k}(\bar{\mA}_{k-1}\bar{ \bm{\Gamma}}_{k-1}\vx_{k}+\hat{\vw}_{k-1})-\vv_{k})$ to the state equation in~\eqref{eq:equiv1}, and obtain the following:
\begin{align*}
    \vx_{k} &= \tilde{\mA}_{k-1}\vx_{k-1}+\tilde{\vu}_{k-1}+ \tilde{\vw}_{k-1},
\end{align*}
where $\tilde{\mA}_{k-1} = (\mI-\mZ_k\mC_{k})\bar{\mA}_{k-1}\bar{ \bm{\Gamma}}_{k-1}$, $\tilde{\vu}_{k-1} \triangleq \mZ_k\vy_{k}$ is the known input, and $\tilde{\vw}_{k-1} \triangleq (\mI-\mZ_k\mC_{k})\hat{\vw}_{k-1}-\mZ_k\vv_{k}$ is the new process noise. The new process noise and the measurement noise could be decoupled by choosing the gain $\mZ_k$ such that
$\mathbb{E}[\tilde{\vw}_{k-1}\vv_{k}^\top]=0$.
The solution can be found by $\mZ_k = \mG_{k-1}\mM_{k}(\mC_{k}\mG_{k-1}\mM_{k})^{-1}$. Then, the system~\eqref{eq:equiv1} becomes 
\begin{align*}
    \vx_{k} &= \tilde{\mA}_{k-1}\vx_{k-1}+\tilde{\vu}_{k-1}+\tilde{\vw}_{k-1}\\
    \vy_{k} &= \mC_{k}\vx_{k}+\vv_{k}.
\end{align*}
Since the pair $(\mC_{k},\tilde{\mA}_{k-1})$ is uniformly detectable, by Theorem 5.2 in~\cite{anderson1981detectability}, the statement holds. 
\hfill $\blacksquare$
\end{thm}

Theorem~\ref{the3} shows that the uniform detectability of the transformed system is one of the sufficient conditions of boundedness of $\mP_k^{x,u}$. Under the assumption of boundedness of $\mP_k^{x,u}$ from Theorem~\ref{the3}, we show the constrained estimation errors $\tilde{\vx}_{k}$ and $\tilde{\vd}_k$ are practically exponentially stable in mean square as in Theorem~\ref{the2}.

\begin{thm} \label{the2}
Consider Assumption~\ref{assum_bounded} and assume that there exist non-negative constants $\ubar{p}$ and $\bar{p}$ such that
$\ubar{p} \mI \leq \mP_k^{x,u} \leq \bar{p} \mI$ holds for all $k$. Then the estimation errors $\tilde{\vx}_{k}$ and $\tilde{\vd}_k$ are practically exponentially stable in mean square, i.e., there exist constants $a_x,a_d,b_x,b_d,c_x,c_d$ such that for all $k$
\begin{align*}
    \mathbb{E}[\|\tilde{\vx}_k\|^2] &\leq a_xe^{-b_xk}\mathbb{E}[\|\tilde{\vx}_0\|^2] +c_x\nonumber\\
    \mathbb{E}[\|\tilde{\vd}_k\|^2] &\leq a_de^{-b_dk}\mathbb{E}[\|\tilde{\vd}_0\|^2] +c_d.
\end{align*}

\noindent {\bf Proof: }
Consider the Lyapunov function
$
    V_k = (\tilde{\vx}^u_{k})^\top (\mP_k^{x,u})^{-1}(\tilde{\vx}^u_{k}).
$
After substituting  \eqref{eq:new unc_state error} into the Lyapunov function, we obtain
\begin{align}\label{lyap_vk}
    V_k =& (\tilde{\vx}^u_{k-1})^\top  \bm{\Gamma}_{k-1}^\top\bar{\mA}_{k-1}^\top(\mI-\mL_k\mC_{k})^\top(\mP_k^{x,u})^{-1}\nonumber\\
    & \times (\mI-\mL_k\mC_{k})\bar{\mA}_{k-1}\bm{\Gamma}_{k-1}\tilde{\vx}^u_{k-1}\nonumber\\
    &+2 (\tilde{\vx}^u_{k-1})^\top \bm{\Gamma}_{k-1}^\top\bar{\mA}_{k-1}^\top(\mI-\mL_k \mC_{k})^\top\nonumber\\
    &\times (\mP_k^{x,u})^{-1}(\mI-\mL_k\mC_{k})\bar{\vw}_{k-1}\nonumber\\
    &+ 2(\tilde{\vx}^u_{k-1})^\top \bm{\Gamma}_{k-1}^\top\bar{\mA}_{k-1}^\top(\mI-\mL_k\mC_{k})^\top(\mP_k^{x,u})^{-1}\bar{\mL}_{k}\vv_{k}\nonumber\\
    &+ \bar{\vw}_{k-1}^\top(\mI-\mL_k \mC_{k})^\top(\mP_k^{x,u})^{-1}(\mI-\mL_k\mC_{k})\bar{\vw}_{k-1}\nonumber\\
    &+ 2\vw_{k-1}^\top(\mI-\mL_k\mC_{k})^\top(\mP_k^{x,u})^{-1}\bar{\mL}_{k}\vv_{k}\nonumber\\
    &+ \vv_{k}^\top \bar{\mL}_{k}(\mP_k^{x,u})^{-1}\bar{\mL}_{k}\vv_{k}.
\end{align}
By the uncorrelatedness property~\cite{papoulis2002probability} of $\vw_{k-1}$, $\vv_k$ and $\tilde{\vx}^u_{k-1}$, the Lyapunov function \eqref{lyap_vk} becomes
\begin{align}
    \mathbb{E}[V_k] &= \mathbb{E}[(\tilde{\vx}_{k-1}^u)^\top \bm{\Gamma}_{k-1}^ \top\bar{\mA}_{k-1}^\top(\mI-\mL_k \mC_{k})^\top(\mP_k^{x,u})^{-1}\nonumber\\
    &\times \bar{\mA}_{k-1}(\mI-\mL_k \mC_{k}) \bm{\Gamma}_{k-1}(\tilde{\vx}_{k-1}^u)]\nonumber\\
    &+ \mathbb{E}[\bar{\vw}_{k-1}^\top(\mI-\mL_k \mC_{k})^\top(\mP_k^{x,u})^{-1}(\mI-\mL_k \mC_{k})\bar{\vw}_{k-1}]\nonumber\\
    &+ \mathbb{E}[\vv_{k}^\top \bar{\mL}_{k}(\mP_k^{x,u})^{-1}\bar{\mL}_{k}\vv_{k}].
    \label{Lyap:exp}
\end{align}
The following statements are formulated to deal with each term in~\eqref{Lyap:exp}.
\begin{claim} \label{cl3}
    There exists a constant $\delta \triangleq (\frac{\ubar{q}'}{\bar{a}'^2\bar{p}}+1)^{-1} \in (0,1)$, such that $ \bm{\Gamma}_{k-1}^\top\bar{\mA}_{k-1}^\top(\mI-\mL_k \mC_{k})^\top (\mP_k^{x,u})^{-1}(\mI-\mL_k \mC_{k})\bar{\mA}_{k-1} \bm{\Gamma}_{k-1} < \delta (\mP_{k-1}^{x,u})^{-1}$.
    
\noindent \textit{Proof: }   
    Since $\rank(\mathcal{B}_k)<n$ $\forall k$, it holds that $\rank(\bar{\mathcal{B}}_k)<n$ $\forall k$ and thus $\bar{ \bm{\Gamma}} \neq 0$. Therefore, $\|\bar{ \bm{\Gamma}}_{k-1}\| = 1$ because $\bm{\gamma}_{k-1}^x\bar{\mathcal{B}}_{k-1}$ is a projection matrix.
    From Assumption \ref{assum_bounded} and Theorem \ref{the3}, we have
        $\bar{\mQ}_{k-1} \geq \ubar{q}'\mI$, and
        $\mP_{k-1}^{x} \leq \bar{p}\mI$.
    Since $\|\bar{\mA}_{k-1}\|$ is upper bounded by $\bar{a}' \triangleq \bar{a}(1+\bar{g}\bar{m}\bar{c}_y)$, we can have $\bar{\mA}_{k-1}\bar{\mA}_{k-1}^\top \leq \bar{a}'^2\mI$. Then we have
    \begin{align}
        \bar{\mQ}_{k-1} &\geq \ubar{q}' \frac{\bar{\mA}_{k-1}\bar{\mA}_{k-1}^\top}{\bar{a}'^2}
        \geq  \frac{\ubar{q}'}{\bar{a}'^2}\bar{\mA}_{k-1}\bar{ \bm{\Gamma}}_{k-1}\bar{ \bm{\Gamma}}_{k-1}^ \top\bar{\mA}_{k-1}^\top\nonumber\\
        &\geq \frac{\ubar{q}'}{\bar{a}'^2\bar{p}} \bar{\mA}_{k-1}\bar{ \bm{\Gamma}}_{k-1} \mP_{k-1}^{x,u} \bar{ \bm{\Gamma}}_{k-1}^ \top\bar{\mA}_{k-1}^\top.
        \label{eq:qbar}
    \end{align}
    Substitution of~\eqref{eq:qbar} into \eqref{eq: new P (Gamma)} yields
    \begin{align}\label{eq:ineq_in_proof}
         &\mP_k^{x,u}- (1+\frac{\ubar{q}'}{\bar{a}'^2\bar{p}})(\mI-\mL_k \mC_{k})\bar{\mA}_{k-1}\bar{ \bm{\Gamma}}_{k-1}\mP_{k-1}^{x,u}\bar{ \bm{\Gamma}}_{k-1}^\top\bar{\mA}_{k-1}^\top \nonumber\\
        &\times (\mI-\mL_k \mC_{k})^\top>0,
    \end{align}
    where the inequality holds because $\mR_k>0$.
     As $(1+\frac{\ubar{q}'}{\bar{a}'^2\bar{p}})\mP_{k-1}^{x,u} >0$, the inverse of the left hand side of \eqref{eq:ineq_in_proof} exists and is symmetric positive definite. By the matrix inversion lemma~\cite{tylavsky1986generalization}, it follows that
    \begin{align}
        &(1+\frac{\ubar{q}'}{\bar{a}'^2\bar{p}})^{-1} (\mP_{k-1}^{x,u})^{-1} - \bar{ \bm{\Gamma}}_{k-1}^ \top\bar{\mA}_{k-1}^\top(\mI-\mL_k \mC_{k})^\top \nonumber \\
        &\times (\mP_k^{x,u})^{-1} (\mI-\mL_k \mC_{k})\bar{\mA}_{k-1}\bar{ \bm{\Gamma}}_{k-1} >0.
        \label{cleq:000}
    \end{align}
    Since $\bm{\gamma}_{k-1}^x\bar{\mathcal{B}}_{k-1}$ is a positive definite matrix, and $\|\bm{\alpha}_{k-1}\|\leq 1$, we have
    \begin{align*}
        \mI-\bm{\gamma}_{k-1}^x\bar{\mathcal{B}}_{k-1} &\leq \bm{\Gamma}_{k-1} \\
        &=\mI-\bm{\gamma}_{k-1}^x\bar{\mathcal{B}}_{k-1}+\bm{\alpha}_{k-1}\bm{\gamma}_{k-1}^x\bar{\mathcal{B}}_{k-1}\leq \mI,
    \end{align*}
    which implies $\| \bm{\Gamma}_{k-1}\|\leq1$.
    Since $\|\bar{ \bm{\Gamma}}_{k-1}\| = 1$ and $\| \bm{\Gamma}_{k-1}\|\leq1$, inequality~\eqref{cleq:000} proves the claim. \hfill $\square$
\end{claim}

\begin{claim}\label{cl4}
    There exists a positive constant $c \triangleq \bar{p}(1+\bar{l}\bar{c}_y)^2(1+\bar{g}\bar{m}\bar{c}_{2})^2\bar{q}\ \rank(\mQ_{k-1}) + \bar{p}(\bar{l}\bar{c}_y\bar{g}\bar{m}-\bar{l}-\bar{g}\bar{m})^2\bar{r_2}\ \rank(\mR_{k})$, such that
    \begin{align*}
        \mathbb{E}[\|(\mI-\mL_k\mC_{k})^\top(\mP_k^{x,u})^{-1}(\mI-\mL_k \mC_{k})\| \|\bar{\vw}_{k-1}\|^2]\\
        +\mathbb{E}[\|\bar{\mL}_{k}(\mP_k^{x,u})^{-1}\bar{\mL}_{k}\| \|\vv_{k}]\|^2]\leq c.\\
    \end{align*}
\noindent \textit{Proof:}
        The first term is bounded by
        \begin{align*}
            \mathbb{E}&[\|(\mI-\mL_k\mC_{k})^\top(\mP_k^{x,u})^{-1}(\mI-\mL_k \mC_{k})\| \|\bar{\vw}_{k-1}\|^2]\\
            =&\mathbb{E}[\|(\mI-\mL_k\mC_{k})^\top(\mP_k^{x,u})^{-1}(\mI-\mL_k \mC_{k})\|\\
            &\|(\mI- \mG_{k-1}\mM_{k}\mC_{k})\|^2\|\vw_{k-1}\|^2]\\
            &\leq \bar{p}(1+\bar{l}\bar{c}_y)^2(1+\bar{g}\bar{m}\bar{c}_{2})^2\bar{q}\ \rank(\mQ_{k-1}),
              \end{align*}
              where we apply $\|\vw_{k-1}\|^2 = \trace(\vw_{k-1}\vw_{k-1}^\top)\leq \bar{q}\ \rank(\mQ_{k-1})$.
      Likewise, the second term is bounded by
              \begin{align*}
             \mathbb{E}&[\|\bar{\mL}_{k}(\mP_k^{x,u})^{-1}\bar{\mL}_{k}\| \|\vv_{k}]\|^2] \\
             &\leq \bar{p}(\bar{l}\bar{c}_y\bar{g}\bar{m}+\bar{l}+\bar{g}\bar{m})^2\bar{r_2}\ \rank(\mR_{k}).
        		\end{align*}
        		These complete the proof. \hfill $\square$
   \end{claim}

Through Claims~\ref{cl3} and~\ref{cl4},~\eqref{Lyap:exp} becomes
\begin{align*}
    \mathbb{E}[V_k] &\leq \delta \mathbb{E}[V_{k-1}] + c.
\end{align*}
By recursively applying the above relation, we have
\begin{align*}
    \mathbb{E}[V_k] &\leq \delta^k\mathbb{E}[V_0] + \sum_{i=0}^{k-1}\delta^i c \leq \delta^k\mathbb{E}[V_0] + \sum_{i=0}^{\infty}\delta^i c\nonumber\\
    & = \delta^k\mathbb{E}[V_0] + \frac{c}{1-\delta},
\end{align*}
which implies practical exponential stability of the estimation error
\begin{align*}
    \mathbb{E}[\|\tilde{\vx}_k^u\|^2] &\leq \frac{\bar{p}}{\ubar{p}}\delta^k\mathbb{E}[\|\tilde{\vx}_0^u\|^2]  +\frac{c\bar{p}}{(1-\delta)}\nonumber\\
    &= a_x'e^{-b_x'k}\mathbb{E}[\|\tilde{\vx}_0^u\|^2] +c_x',
\end{align*}
where $(\tilde{\vx}_{k}^u)^\top (\mP_k^x)^{-1} (\tilde{\vx}_{k}^u) \geq \lambda_{\min}{((\mP_k^x)^{-1})}\|\tilde{\vx}_{k}^u\|^2 \geq \frac{1}{\bar{p}}\|\tilde{\vx}_{k}^u\|^2$
and
$(\tilde{\vx}_{0}^u)^\top (\mP_0^x)^{-1} \tilde{\vx}_{0}^u \leq \lambda_{\max}{((\mP_0^x)^{-1})}\|\tilde{\vx}_{0}^u\|^2 \leq \frac{1}{\ubar{p}}\|\tilde{\vx}_{0}^u\|^2$ have been applied.
Constants are defined by
\begin{align*}
    &a_x'\triangleq \frac{\bar{p}}{\ubar{p}},
    &&b_x'\triangleq \ln(1+\frac{\ubar{q}'}{\bar{h}^2\bar{a}'^2\bar{p}})
    &&c_x'\triangleq \frac{c\bar{p}}{(1-\delta)}.
\end{align*}
Since $\tilde{\vx}_{k}$ is a linear transformation of $\tilde{\vx}_{k}^u$, the same stability holds for $\tilde{\vx}_{k}$.
Likewise, the same stability holds for $\tilde{\vd}_k$ in~\eqref{eq:dtile} because it is a linear transformation of $\tilde{\vx}_{k}$. We omit its details.
\hfill $\blacksquare$
\end{thm}

\section{Illustrative Example} \label{simulation}

In this example, we test Algorithm~\ref{algorithm1} on a vehicle model with input and state constraints and compare the estimation accuracy and the detection performance with an unconstrained algorithm. 

\begin{table*}[ht] 
\begin{center}
\caption{Performance comparison.}
\label{table: sum}
 \begin{tabular}{m{2.5cm} m{3.cm} m{3.cm} m{3.cm} m{3.cm}} 
 \hline
  & $\sum_k\|\tilde{\vx}_k\|$ & $\sum_k\|\tilde{\vd}_k\|$& $\sum_k\| \trace(\mP^{x}_k)\|$ &   $\sum_k\|\trace(\mP^d_k)\|$  \\ [0.5ex] 
 \hline\hline
 \texttt{CARE} & 88.928 & 672.914 & 0.455 & 27.351$^*$ \\ 
 \hline
 \texttt{ISE}& 123.623 & 1041.837 & 0.613 & 40.577$^*$  \\[.5ex] 
 \hline
\end{tabular}
\end{center}
\vspace{-2mm}
\footnotesize{\it $^*$ The summation ranges from $k=100$ to $k=1000$ due to the large initialization ($10^4$-scale), as shown in Fig.~\ref{fig: attack_trace}.}
\end{table*}

\subsection{Experimental Setup}
\begin{figure}[ht]
\begin{center}
\includegraphics[height=3cm]{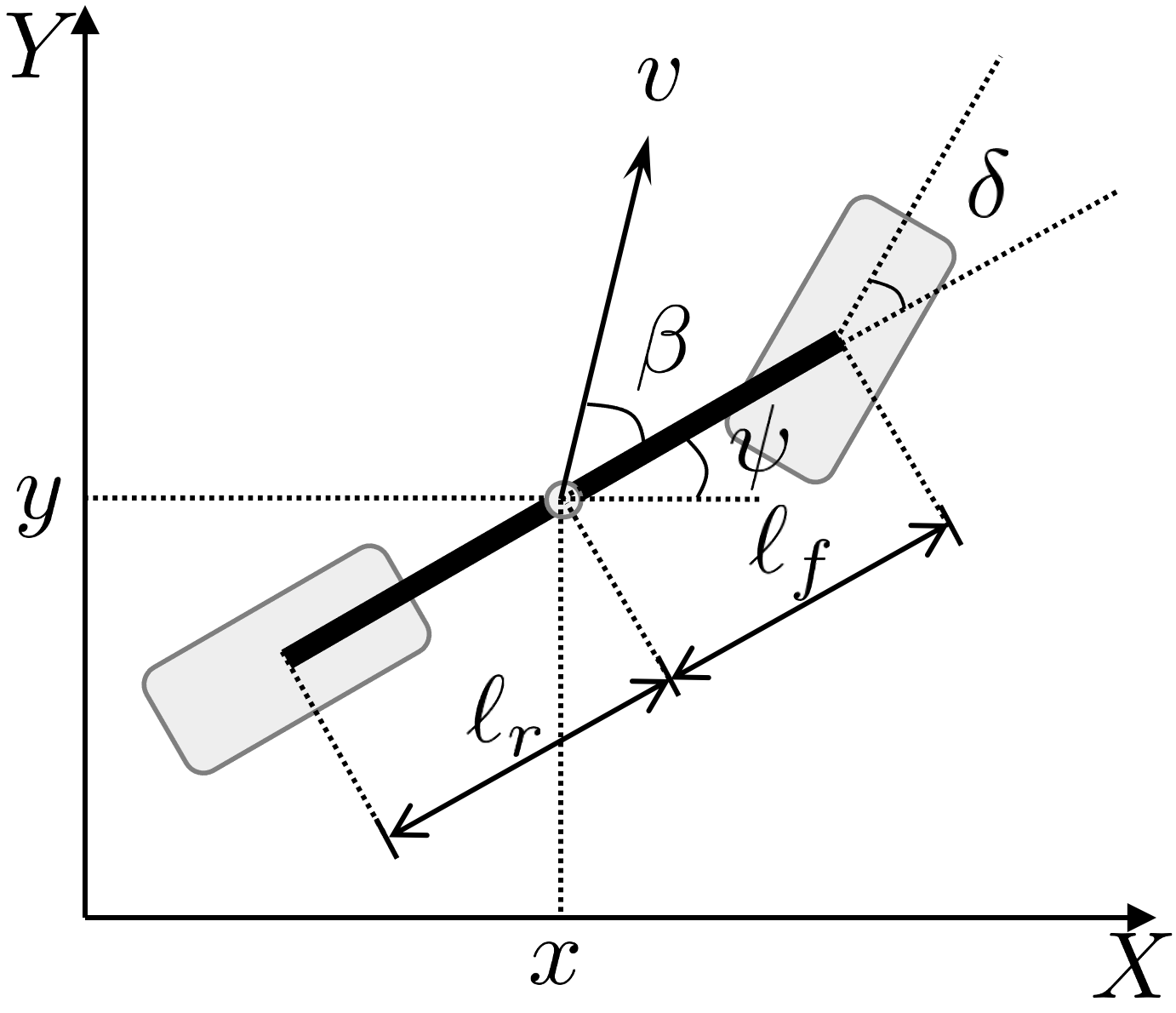} 
\caption{Kinematic Bicycle Model.}
\label{fig: vd}                            
\end{center} 
\vspace{-8mm}
\end{figure}
We consider a kinematic bicycle model (Fig.~\ref{fig: vd}) in~\cite{rajamani2011vehicle}. The nonlinear continuous-time model is given as
\begin{align*}
    \dot{x} &= v \cos(\psi + \beta) \\
    \dot{y} &= v \sin(\psi + \beta) \\
    \dot{\psi} &= \frac{v}{l_r} \sin(\beta) \\
    \dot{v} &= a \\
    \beta &= \arctan \Big(\frac{l_r}{l_f+l_r} \tan(\delta) \Big),
\end{align*}
where $x$ and $y$ are the coordinates of the center of mass, $v$ is the velocity of the center of mass, $\beta$ is the angle of the velocity $v$ with respect to the longitudinal axis of the vehicle, $a$ is the acceleration, $\psi$ is the heading angle of the vehicle, $\delta$ is the steering angle of the front wheel, and $l_f$ and $l_r$ represent the distance from the center of mass of the vehicle to the front and rear axles, respectively.

Since the proposed algorithm is for linear discrete-time systems, we perform the linearization and discretization as in~\cite{law2018robust} with sampling time $T_s = 0.01s$. We rewrite the system in the form of~\eqref{eq: sys}, where $\vx_k = [x_k, y_k, \psi_k, v_k]^\top$ is the state vector, $\vu_k =  [\beta_k^u, a_k^u]^\top = \Big[\arctan \Big(\frac{l_r}{l_f+l_r} \tan(\delta_k^u) \Big), a_k^u\Big]^\top$ is the input vector, and $\vd_k =  [\beta_k^d, a_k^d]^\top = \Big[\arctan \Big(\frac{l_r}{l_f+l_r} \tan(\delta_k^d) \Big), a_k^d\Big]^\top$ is the attack input vector. We consider the scenario that attack input is injected into the input, i.e. $\mG_k = \mB_k$. The system matrices are given as follows:
\begin{align*}
    \mA_k = 
    \begin{bmatrix}
    1 & 0 & 0 & T_s\\
    0 & 1 & v_k T_s & 0 \\
    0 & 0 & 1 & 0 \\
    0 & 0 & 0 & 1
    \end{bmatrix}, \
    \mB_k = \mG_k =
    \begin{bmatrix}
    0 & 0 \\ 
    v_k T_s & 0\\
    \frac{v_k T_s}{l_r}  & 0\\
    0 & T_s
    \end{bmatrix},
\end{align*}
and $\mC_k = \mI$.
The noise covariances $\mQ_k$ and $\mR_k$ are considered as diagonal matrices with $\diag(\mQ_k) = [0.1, 0.1, 0.001, 0.0001]$ and $\diag(\mR_k) = [0.01, 0.01, 0.001, 0.00001]$.

The vehicle is assumed to have state constraints on the location $0\leq x_k \leq 20$, $0\leq y_k \leq 5$ and the velocity $0\leq v_k \leq 22$, and input constraints on the steering angle $|\delta| \leq 1.0472$ and the acceleration $|a| \leq 3.5$. 

The unknown attack signals are
\begin{align*}
    \delta_k^d &=
    \begin{cases}
    0, & 0\leq k < 100\\
    1.1  \sin(0.05k), & 0\leq k < 100
    \end{cases} \\
    a_k^d &=
    \begin{cases}
    0 & 0\leq k < 100 \\
    3.5, &  100 n \leq k < 100(n+1)\\
    - 3.5, &  100(n+1) \leq k < 100(n+2)
    \end{cases},
\end{align*}
where $n = 1,2,\cdots, 5$.

The constraints on the vehicle can be formulated by inequality constraints as in~\eqref{eq: const}:
\begin{align*}
\underbrace{
    \begin{bmatrix}
    1 & 0 \\
    -1 & 0 \\
    0 & 1 \\
    0 & -1
    \end{bmatrix}
}_{\mathcal{A}_{k-1}}
    \begin{bmatrix}
    \delta_{k-1}^d\\
    a_{k-1}^d
    \end{bmatrix}
    &\leq
\underbrace{
    \begin{bmatrix}
    1.0472 - \delta_{k-1}^u\\
    1.0472 + \delta_{k-1}^u\\
    3.5 - a_{k-1}^u\\
    3.5 + a_{k-1}^u
    \end{bmatrix}
}_{\vb_{k-1}} \\
\underbrace{
    \begin{bmatrix}
    1 & 0 & 0 & 0\\
    -1 & 0 & 0 & 0 \\
    0 & 1 & 0 & 0 \\
    0 & -1 & 0 & 0 \\
    0 & 0 & 0 & 1 \\
    0 & 0 & 0 & -1
    \end{bmatrix}
}_{\mathcal{B}_{k}}
    \begin{bmatrix}
    x_k\\
    y_k\\
    \psi_k\\
    v_k
    \end{bmatrix}
    &\leq
\underbrace{
    \begin{bmatrix}
    20\\
    0\\
    5\\
    0 \\
    22 \\
    0
    \end{bmatrix}
}_{\vc_{k}}
.
\end{align*}
To reduce the effect of instantaneous noises, the cumulative sum algorithm (CUSUM) is adopted~\cite{lai1995sequential}. The $\chi^2$ test is utilized in a cumulative form. The $\chi^2$ CUSUM detector is characterized by the detector state $S_k\in\mathbb{R}_{+}$:
\begin{align}
    S_{k}=\phi S_{k-1}+\hat{\bm{d}}_k^\top \bm{P}_k^{-1}\hat{\bm{d}}_k, \quad S_0=0,
    \label{eq: detector state}
\end{align}
where $0<\phi<1$ is the pre-determined forgetting rate. At each time $k$, the CUSUM detector~\eqref{eq: detector state} is used to update the detector state $S_k$ and detect the attack. In particular, we conclude that the attack is presented if
\begin{align}
     S_k>\sum_{i=0}^{\infty}\phi^i\chi^2_{df}(\alpha)=\frac{\chi^2_{df}(\alpha)}{1-\phi}. \label{eq: cum threshold}
\end{align}
All values are in standard SI units: $m$ (meter) for $l_f$, $l_r$, $x_k$, and $y_k$; $rad$ for $\delta^u_k$, $\delta^d_k$, $\beta^u_k$, $\beta^d_k$, and $\psi_k$; $m/s$ for $v_k$; $m/s^2$ for $a^u_k$ and $a^d_k$.
\begin{figure}[!ht]
\begin{center}
\includegraphics[width=6.7cm]{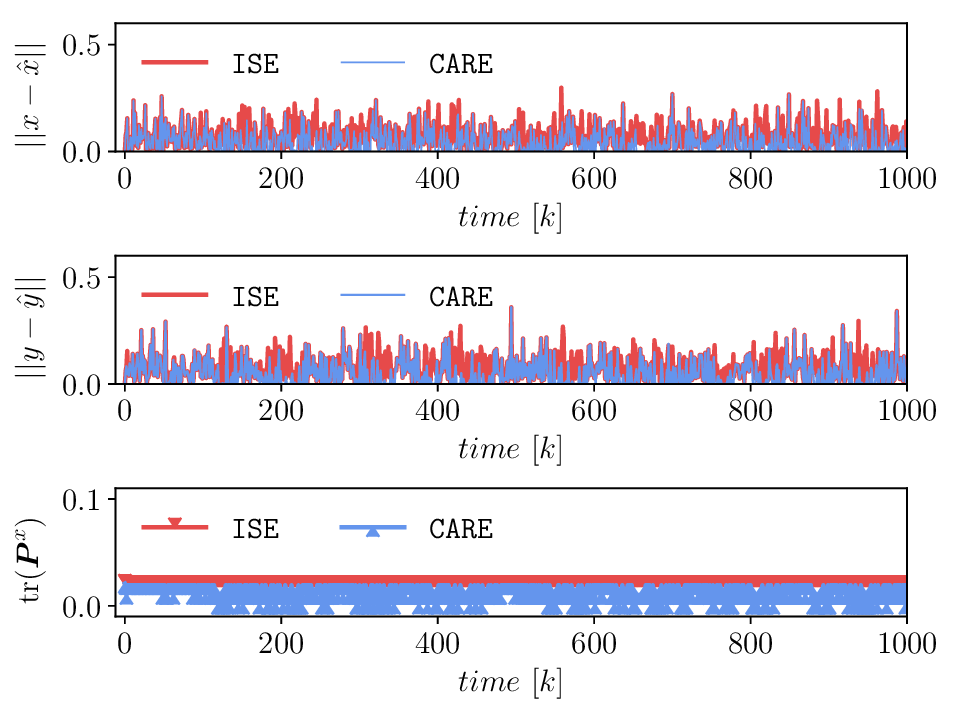}
\caption{Estimation errors of constrained states and traces of the state error covariance.} 
\label{fig: state_trace}
\vspace{-9mm}
\end{center}        
\end{figure}
\subsection{Results}
\begin{figure}[!ht]
\begin{center}
\includegraphics[width=6.7cm]{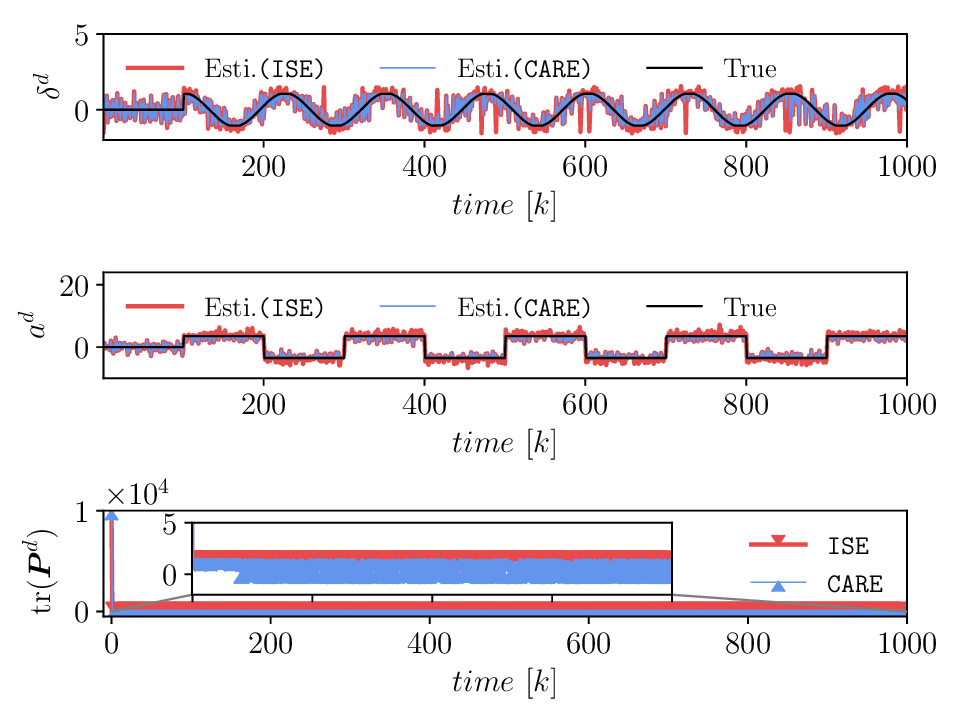}
\caption{Attack signal estimation and traces of error covariance of the attack signals.}
\label{fig: attack_trace}
\vspace{-7mm}
\end{center}          
\end{figure}

\begin{figure}[!ht]
\begin{center}
\includegraphics[width=6.7cm]{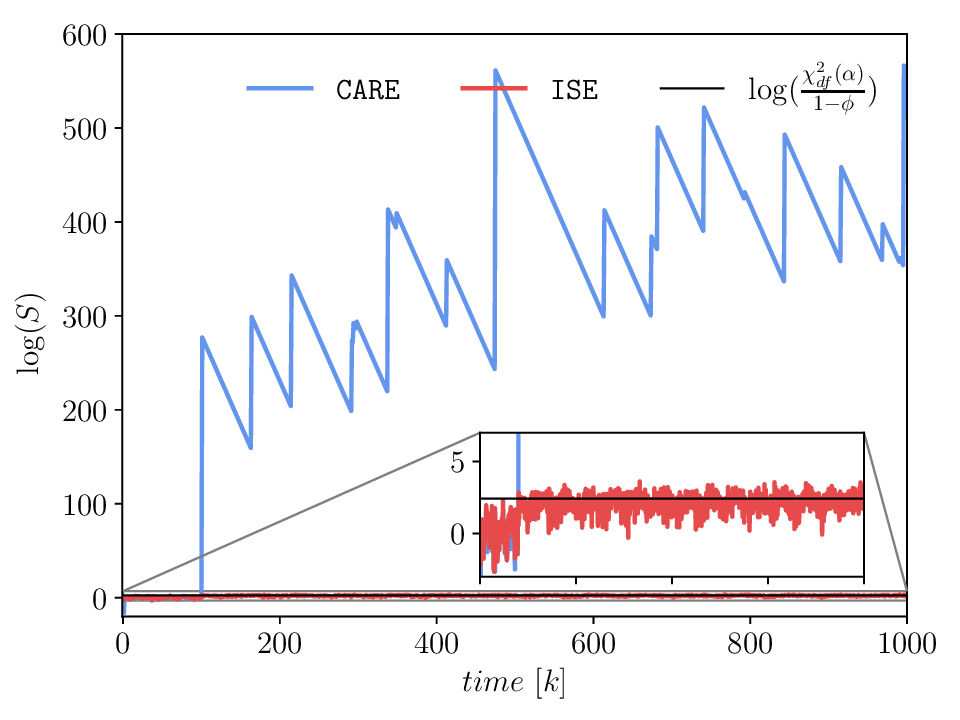}
\caption{Attack detection.}
\label{fig: detection state}
\vspace{-8mm}
\end{center}    
\end{figure}

We show a comparison of the proposed algorithm (\texttt{CARE}) and the unified linear input and state estimator (\texttt{ISE}) introduced in~\cite{yong2016unified}. Figure~\ref{fig: state_trace} shows the estimation errors of the constrained states ($x_k$ and $y_k$) and the traces of the state error covariances, and Fig.~\ref{fig: attack_trace} shows the unknown attack signals and their estimates and traces of the attack estimation error covariances. As expected, \texttt{CARE} produces smaller state estimation error and lower covariance. When the attack happens after $k=100$, the estimates obtained by \texttt{CARE} are closer to the true values and have lower error covariances (cf. Table~\ref{table: sum}).
The estimates are used to calculate the detector state $S_k$ in~\eqref{eq: detector state}. The statistical significance of the attack is tested using the CUSUM detector. The threshold is calculated by $\chi^2_{df} / (1-\phi)$ in~\eqref{eq: cum threshold} with the significance level $\alpha = 0.01$ and the forgetting rate $\phi = 0.15$. The detector states and the threshold are plotted in $log-$scale (Fig.~\ref{fig: detection state}). When the attack is present, \texttt{CARE} can detect the attack by producing high detector state values above the threshold, while the detector state values from \texttt{ISE} are oscillating around the threshold, suffering from a high false negative rate of $66.44\%$.
\section{Conclusion} \label{conclusion}
In this paper, we presented a constrained attack-resilient estimation algorithm (\texttt{CARE}) of linear stochastic cyber-physical systems. The proposed algorithm produces minimum-variance unbiased estimates and utilizes physical constraints and operational limitations to improve estimation accuracy and detection performance via projection. In particular, \texttt{CARE} first provides minimum-variance unbiased estimates, and then these estimates are projected onto the constrained space. We formally proved that estimation errors and their covariances from \texttt{CARE} are less than those from unconstrained algorithms and showed that \texttt{CARE} had improved false negative rate in attack detection. Moreover, we proved that the estimation errors of the proposed estimation algorithm are practically exponentially stable. A simulation of attacks on a vehicle demonstrates the effectiveness of the proposed algorithm and reveals better attack-resilient properties compared to an existing algorithm.

\appendix
\section{Gauss-Markov Theorem} \label{Thm:GMThm}
\begin{thm}[Gauss-Markov Theorem~\cite{sayed2003fundamentals}]
Given the linear model $\vy = \mH \vx +\vv$, where $\vv$ is a zero-mean random variable with positive-definite covariance matrix $\mR_v$ and $\mH$ is full rank $m \times n$ matrix with $m \geq n$, the minimum-variance-unbiased linear estimator of $\vx$ given $\vy$ is $\hat{\vx} = (\mH^{*} \mR_v^{-1} \mH)^{-1}\mH^{*}\mR_v^{-1} \vy$.
\end{thm}

\section*{Acknowledgements}
This work has been supported by the National Science Foundation (ECCS-1739732 and CMMI-1663460).
\bibliography{references}
\bibliographystyle{elsarticle-num}
\end{document}